\documentclass[12pt]{article}
\usepackage{amsmath}
\usepackage{amssymb}
\usepackage{psfig}
\usepackage{graphicx,color}

\numberwithin{equation}{section}

\newcommand{\qed}{\hfill \mbox{\raggedright \rule{.07in}{.1in}}}

\newcommand{\R}{{\mathbb R}}

\renewcommand{\tilde}{\widetilde}

\newcommand{\To}{\rightarrow}

\title{Testing for Chaos in Deterministic \\ Systems with Noise}
\author{
Georg A. Gottwald  % \thanks{} 
\\ School of Maths and Stats
\\ University of Sydney
\\ NSW 2006, Australia
\and
Ian Melbourne 
\\ Department of Maths and Stats
\\ University of Surrey
\\ Guildford GU2 7XH, UK
}

\date{12 October, 2004}

\begin{document}

\maketitle

\begin{abstract}
Recently, we introduced a new test for distinguishing regular from
chaotic dynamics in deterministic dynamical systems and argued that
the test had certain advantages over the traditional test for chaos
using the maximal Lyapunov exponent.

In this paper, we investigate the capability of the test to cope with
moderate amounts of noisy data.  Comparisons are made between an
improved version of our test and both the ``tangent space'' and
``direct method'' for computing the maximal Lyapunov exponent.  The
evidence of numerical experiments, ranging from the logistic map to an
eight-dimensional Lorenz system of differential equations (the Lorenz
96 system), suggests that our method is superior to tangent space
methods and that it compares very favourably with direct methods.
\end{abstract}


% \newpage

\section{Introduction} 
\label{sec-intro}
%% When studying time series with the aim to distinguish between regular
%% and chaotic dynamics in deterministic dynamical systems one has tried
%% to distill quantities which are invariant under measurements and
%% changes in the measurement procedures. Often used quantities are
%% Lyapunov exponents and the correlation sum (see for example

%% Although the entire  spectrum of Lyapunov
%% exponents can be determined in principle,
%% one is often only interested whether a given time series
%% originated from a regular or a chaotic deterministic dynamical
%% system. In such a case it is sufficient to calculate only the maximal
Given time series data from a deterministic dynamical system, it is often of
interest to determine whether the underlying dynamics are regular or chaotic.
For example, heart rate data measured in
electrocardiograms are believed to be chaotic for the case of healthy
patients but show regularity in the case of congestive heart failure
(see Wagner \& Persson \cite{Wagner98} for an overview).

The usual test of whether a deterministic dynamical system is chaotic
or nonchaotic is the calculation of the maximal Lyapunov exponent
$\lambda$. 
Standard references include
\cite{Abarb93,Eckmann86,KantzSchreiber,Lauterborn97,ParkerChua89,Schreiber99}. 
A positive maximal Lyapunov exponent indicates chaos: if
$\lambda>0$, then nearby trajectories separate exponentially and if
$\lambda\le0$, then nearby trajectories stay close to each other. This
approach has been widely used for dynamical systems whose equations
are known. If the
equations are not known or one wishes to examine experimental data,
then $\lambda$
may be estimated using the phase space reconstruction method of Takens
\cite{Takens81} or by approximating the linearisation of the evolution
operator.  
%% Nevertheless, the computation of Lyapunov exponents is
%% greatly facilitated if the underlying equations are known and are
%% low-dimensional.

In a previous paper~\cite{GM04}, we introduced a binary test for
distinguishing between regular and chaotic dynamics in deterministic
dynamical systems.  The test has two main
advantages over computing the maximal Lyapunov exponent:
\begin{itemize}
\item[(i)] The test applies directly to time series data, so that
phase space reconstruction is not required.
Moreover, the form and nature of the underlying dynamical
system is irrelevant; the test applies equally well to continuous time
systems and discrete time systems, to experimental data and maps,
to ordinary differential equation and partial differential equations.
\item[(ii)] It is a binary test (in principle, the test yields $0$ or $1$), so 
a numerically computed value of $0.01$ say yields a definite conclusion,
whereas such a value for the maximal Lyapunov exponent would not.
\end{itemize}

The theoretical basis for the validity of any numerical test for chaos relies
on the availability of unlimited noiseless data. In practice, it is
necessary to work with limited amounts of contaminated data. The only
way to determine the utility of the test in such situations is to try
it out on different examples and see how it fares in comparison with
existing methods. Such a comparison is carried out in this paper.

In Section~\ref{sec-test}, we describe  an improved version of
the test introduced in~\cite{GM04}.
In Section~\ref{sec-tradmeth}, we briefly
review the tangent space and direct methods for computing maximal
Lyapunov exponents. In Section~\ref{sec-logistic}, we
apply the test to the logistic map contaminated with measurement noise. 
Here, our test compares extremely favourably to the tangent space method for
computing the Lyapunov exponent.
In Section~\ref{sec-rosenstein}, the test is applied to contaminated data
from an eight-dimensional ODE and is seen to compare favourably even to 
direct methods.

%%%%%%%%%%%%%%%%%%%%%%%%%%%%%%%%%%%%%%%%%%%%%%%%%%%%%%%%%%%%%%%%%%%%
%%%%%%%%%%%%%%%%%%%%%%%%%%%%%%%%%%%%%%%%%%%%%%%%%%%%%%%%%%%%%%%%%%

\section{The (modified) test for chaos}
\label{sec-test}

Our test for chaos in this paper is a slightly modified version of the
test that we introduced previously~\cite{GM04}.
In this section, we describe the modified test and compare it with the 
one in~\cite{GM04}.
We focus throughout on discrete data sets $\phi(n)$ sampled at times
$n=1,2,3,\ldots$  (The continuous time case is similar, cf.~\cite{GM04}.) 
Here, $\phi(n)$ is a one-dimensional observable obtained from the underlying
dynamics.

Choose a constant $c\in\R$ at random.
In practice, we choose a number
of different values of $c$ as detailed below.
For each value of $c$, define
\begin{align}
\label{modp}
p(n)=\sum_{j=1}^n \phi(j) \cos(jc), \quad n=1,2,3,\ldots
\end{align}
%% It follows from ergodic theoretical ideas (see~\cite{GM04}) that 
%% \begin{itemize}
%% \item[(i)] $p(n)$ is bounded if the underlying dynamics is
%% nonchaotic and 
%% \item[(ii)]
%% $p(n)$ behaves asymptotically like a Brownian motion
%% if the underlying dynamics is chaotic.
%% \end{itemize}
Next, define the mean square displacement 
\[
M(n)=\lim_{N\To\infty}\frac1N\sum_{j=1}^{N}[p(j+n)-p(j)]^2, 
\quad n=1,2,3,\ldots
\]
If the dynamics is regular (periodic or quasiperiodic), then with
probability one $M(n)$ is a bounded function of $n$. However, if the
dynamics is chaotic (in a fairly mild sense), then with probability
one $M(n)=Vn+{\cal{O}}(1)$ for some $V>0$.
(For justifications of these statements, see~\cite{GM04} and the
references therein.)   Define the asymptotic growth rate of the mean square
displacement
\[
K=\lim_{n\To\infty} \frac{\log M(n)}{\log n}.
\]
Then $K=0$ signifies regular dynamics whereas $K=1$ signifies chaotic
dynamics.

Next suppose that we have a finite amount of data
$\phi(n)$, $1\le n\le N$.  Define 
\[
M(n)=\frac{1}{N-n} \sum_{j=1}^{N-n}[p(j+n)-p(j)]^2.
\]
Provided $n\ll N$ it is reasonable to expect that $M(n)$ scales with
$n$ in a similar way to before. To avoid logarithms of negative
numbers, we plot $\log (M(n)+1)$ against $\log n$ for $1\le n\le N_1$
for some choice of $N_1$, $1\ll N_1 \ll N$.
In practice, we have found it convenient to choose $N_1=N/10$.
This choice is made throughout the paper without further comment.
(A disadvantage of our test is that we cannot
use the full available data set for our statistics in the
computation of the mean square displacement.) 
We define $K$ to be the
slope of the line of best fit on the resulting data set, using
least square regression.  The test is
now that $K$ close to zero signifies regular dynamics and $K$ close to
$1$ implies chaotic dynamics.

For short data sets (with or without noise) there is an inconvenient resonance
phenomenon, where a given choice of $c$ may resonate with frequencies
in the underlying dynamics.   For example, if the signal is $2\pi$-periodic
and we choose $c$ to be an integer, then a simple argument using the
Fourier series for $\phi(n)$ shows that typically $p(n)$ will  grow linearly
yielding $K=2$.   A near-integer choice of $c$ results in $K=0$ eventually
but the convergence is very slow; for a small number of iterates the
computed value of $K$ is likely to be closer to $1$ than to $0$.
The situation for quasiperiodic dynamics is even worse
since ``bad'' choices of $c$ are now dense.   Nevertheless,
there is zero probability of making a bad choice in theory.
In practice, numerical experimentation shows that even 
for short data series, the bad choices of $c$ are rare but do occur from 
time-to-time.

Our resolution of this problem is to choose several 
values of $c$ at random, computing $K$ for each choice of $c$. 
Then we take the median value of $K$. (We do not take the average of $K$,
since when $c$ does fail it can fail quite badly.)  

To demonstrate the improvement in the modified test, we consider the
logistic map
\begin{align*}
x_{n+1}=\mu x_n(1-x_n),
%% \label{logi}
\end{align*}
varying the parameter $\mu$ in the range $3.5\le\mu\le 4$
in increments of $0.001$.
Starting with initial condition $0.0001$, we use
$N=1,000$ iterates, after a transient of $20,000$ iterates.
We take $\phi(n)=x_n$.

\begin{figure}[hb]
\centerline{%
% logistic/compxx.slope
\includegraphics[angle=-90,width=0.5\textwidth]{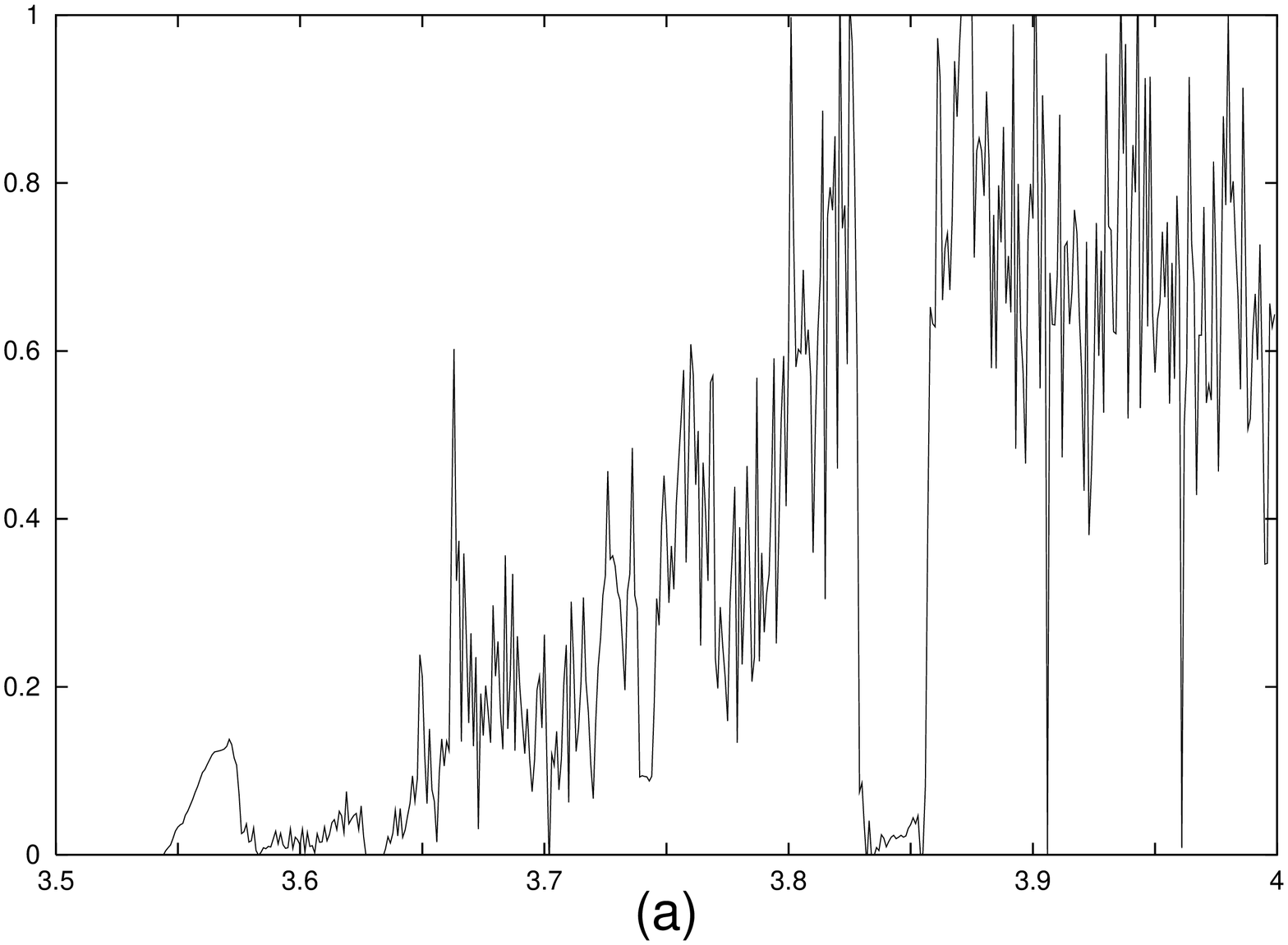}
\includegraphics[angle=-90,width=0.5\textwidth]{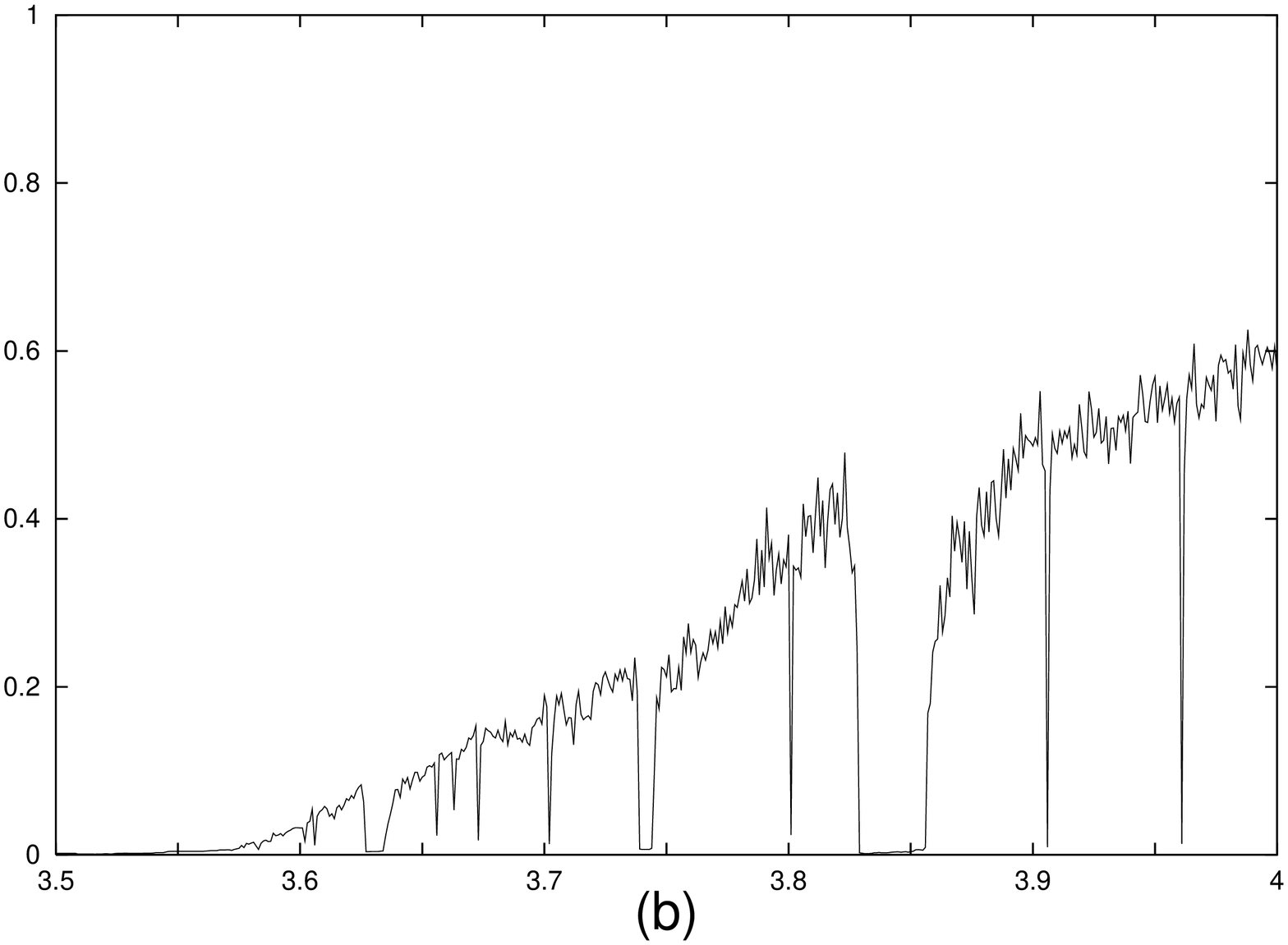}}
%% \centerline{%
%% % logistic/compxx.slope
%% \includegraphics[angle=-90,width=0.5\textwidth]{FIGURES/comp2.ps}
%% \includegraphics[angle=-90,width=0.5\textwidth]{FIGURES/comp2M.ps}}
\caption{\label{fig-oldtest} Plots of 
$K$ versus $\mu$ for the logistic map $f(x)=\mu x(1-x)$,
 with $3.5\le \mu\le 4$ using $1,000$ iterates and
noise-free data. (a) The old test~\cite{GM04}. 
(b)  The new test with $100$ values of $c$.}
 %% In the
%% upper panel $K$ is shown for noise-free data. In the lower panel we
%% show results when measurement noise of $10\%$ is added to the
%% data. The left pictures show the old test, the right pictures the
%% modified test where the median value of $K$ is taken from $100$ 
%% values of $c$.}
\end{figure}

In Figure~\ref{fig-oldtest}, we show how the
old method~\cite{GM04} compares with the modified test, where for the
modified test we
take the median of $100$ different values for $c$. It
is clear that there is a dramatic improvement in our test for chaos.
(A comparison with traditional Lyapunov exponent methods is carried
out in Section~\ref{sec-logistic}.)
An obvious criticism is that $K$ is always far from $1$ for such short data
sets.   In later sections, we demonstrate that this issue is
of little consequence in comparison with traditional methods.
Our purpose in this section is solely to compare the modified test
with our old test.

 \clearpage

\begin{figure}[hb]
\centerline{%
\includegraphics[angle=-90,width=0.5\textwidth]{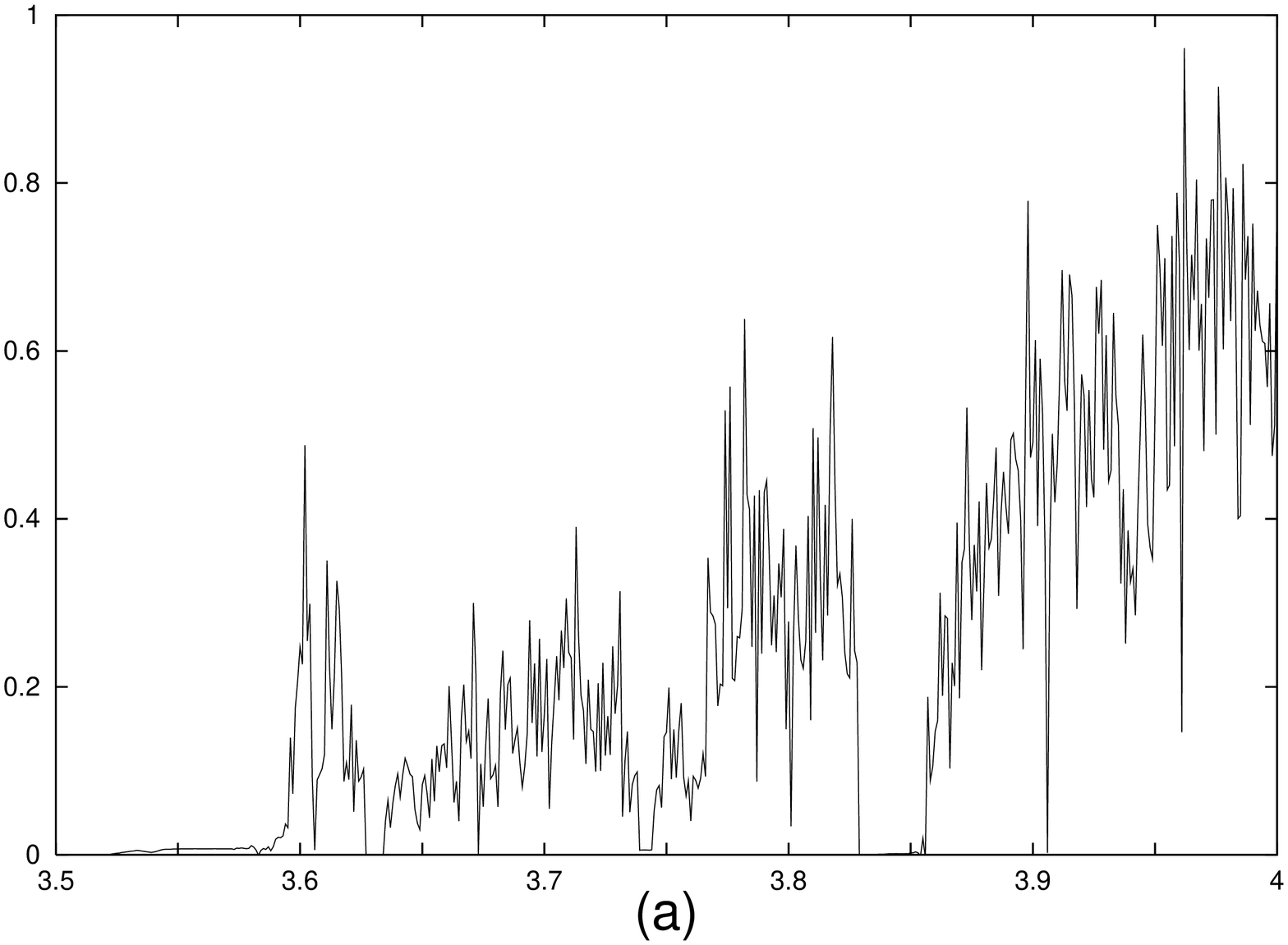}
\includegraphics[angle=-90,width=0.5\textwidth]{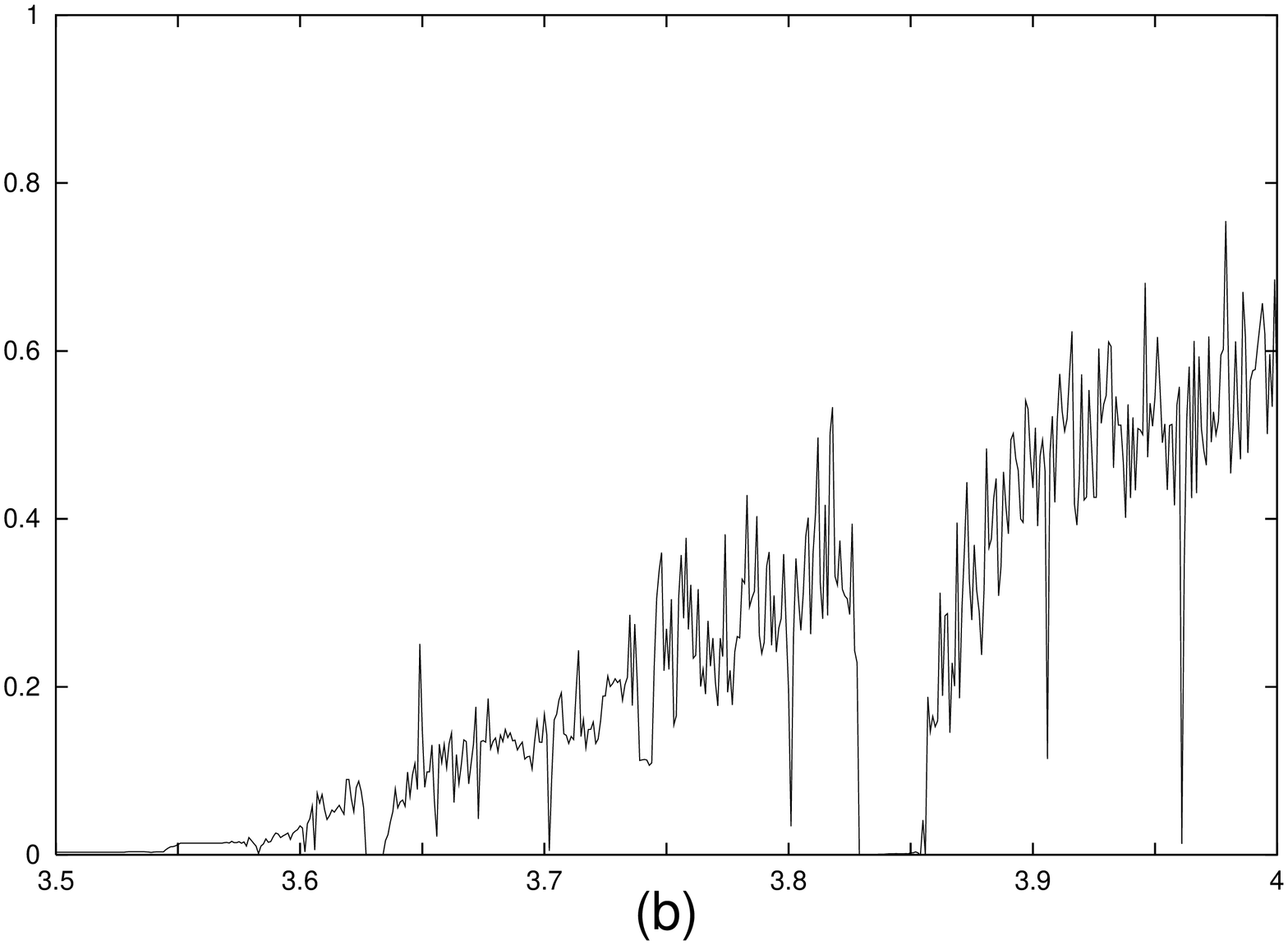}}
\centerline{%
\includegraphics[angle=-90,width=0.5\textwidth]{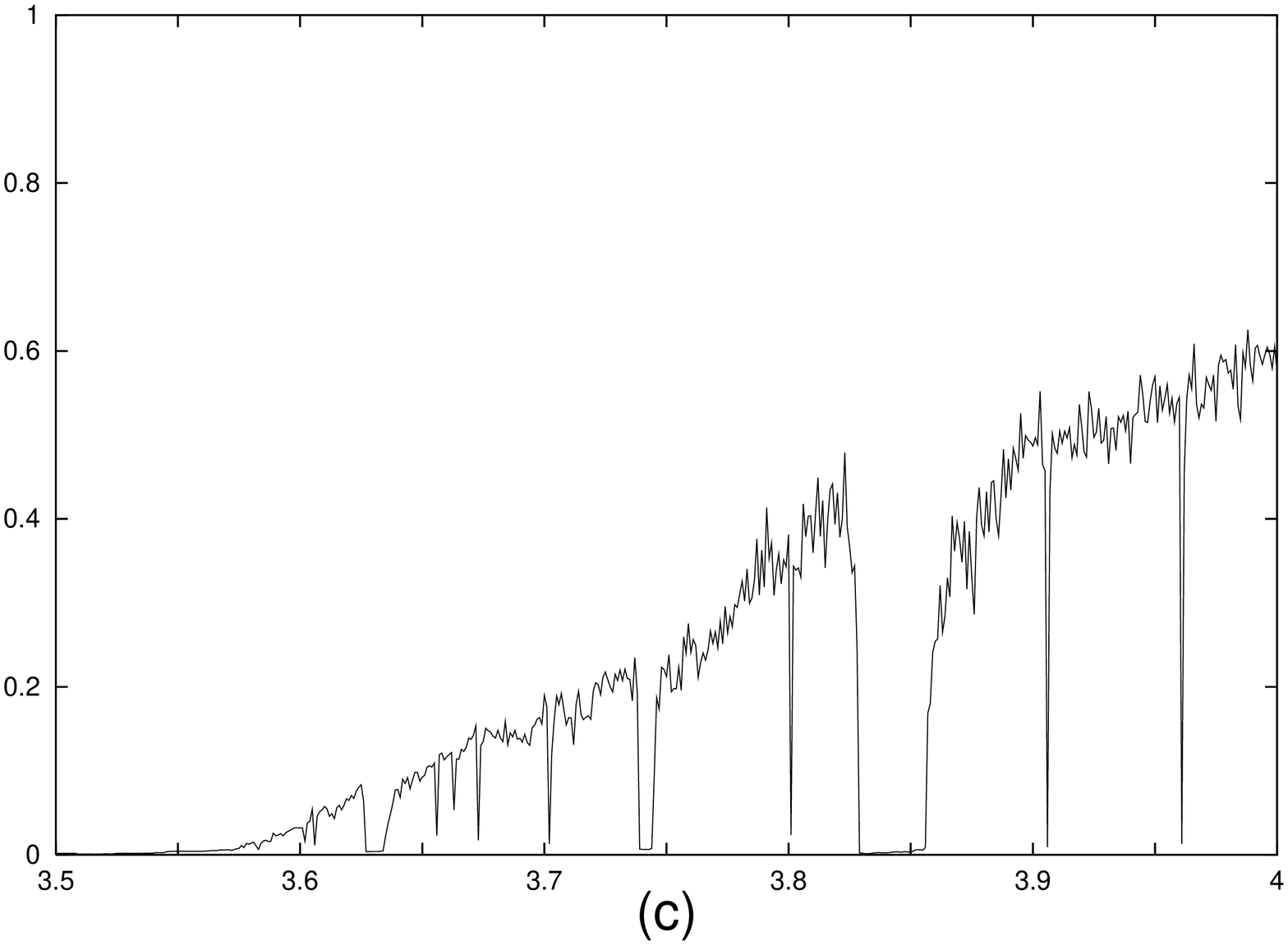}
\includegraphics[angle=-90,width=0.5\textwidth]{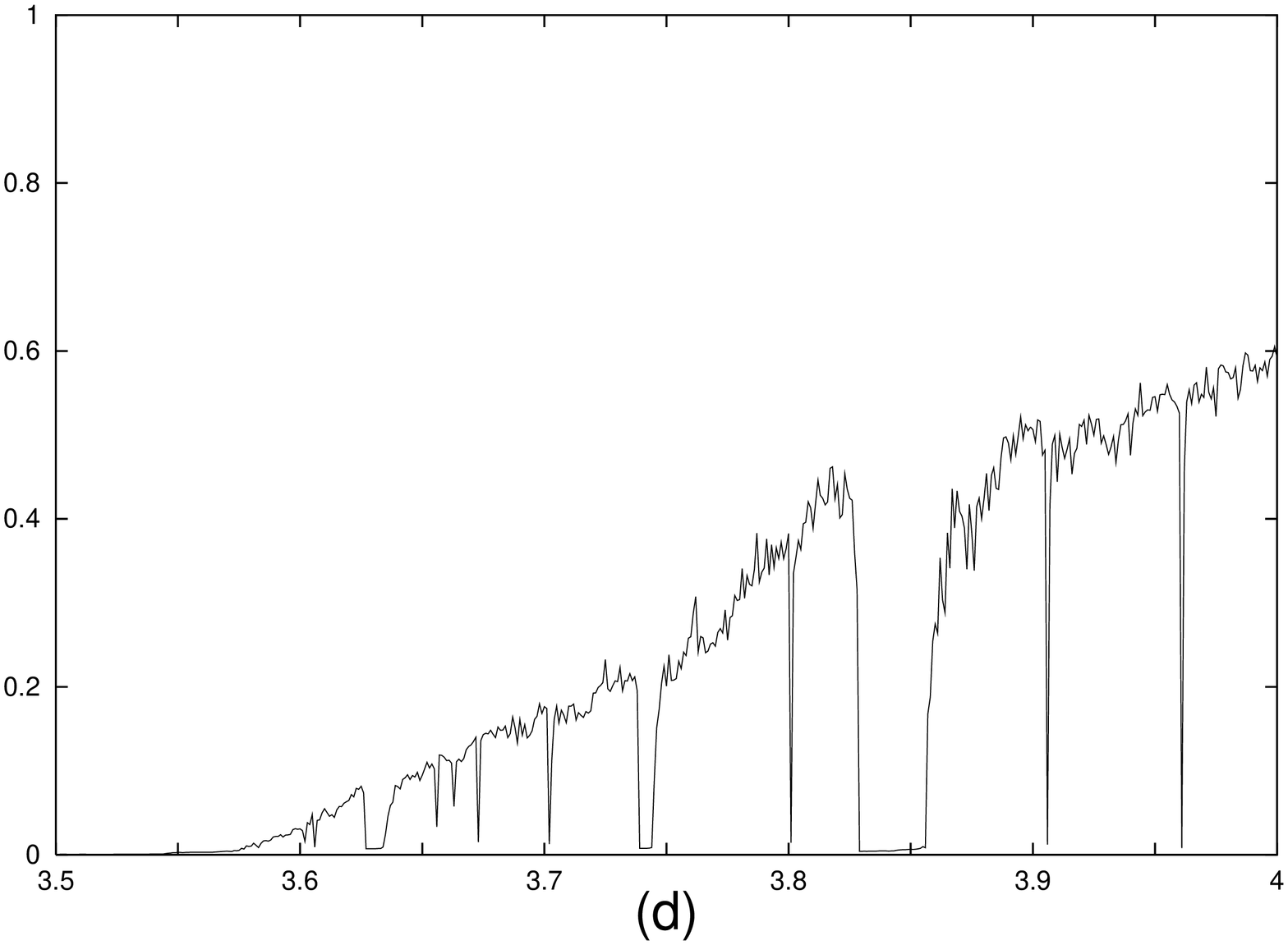}}
\caption{\label{fig-median} Plots of 
$K$ versus $\mu$ for the logistic map $f(x)=\mu x(1-x)$,
 with $3.5\le \mu\le 4$ using $1,000$ iterates and noise-free data.   
The median value of $K$ is taken from $m$ values of $c$
where (a) $m=1$, (b) $m=10$, (c) $m=100$, (d) $m=1000$.}
\end{figure}

In Figure~\ref{fig-median}, we compare the effects of taking $1$, $10$, $100$
and $1000$ different values of $c$.  It is clear from these results
that taking one choice of $c$ is not effective and that taking $10$
values of $c$ is a dramatic improvement, though some periodic windows are still
poorly defined.  This is remedied by going to $100$ values of $c$, and
increasing to $1000$ does not make a significant further improvement.
In the remainder of the paper, we use $100$ values of $c$.

%% When there is noise present in the data $\phi(n)$, the test fails if
%% there is too much data (in the same way that Lyapunov exponents fail
%% -- eventually the noise dominates). As with Lyapunov exponents, the
%% issue is whether we can simultaneously address convergence (which
%% requires large $N$) and noise (which requires small $N$). The success
%% of this balancing act can only be judged by appealing to a higher
%% authority, or looking at examples.  In the subsequent sections, we
%% look at a variety of different situations and check how our method
%% compares with the Lyapunov exponent method.
 
%\begin{rmk}
{\bf Remark 2.1} There are two differences between the modified test
presented here and the test introduced in~\cite{GM04}.  The first is
that we now use several values of $c$.  The second is that
in~\cite{GM04} we defined $p(n)$ by the more complicated prescription:
\begin{align} 
\begin{array}{l} \theta(n+1) = \theta(n) + c + \phi(n) \\ 
p(n+1) = p(n) + \phi(n) \cos(\theta(n))
\end{array} \biggr\}.
\label{thetap}
\end{align}
We note that the modified $p(n)$ in equation~\eqref{modp} can be recovered
by removing the $\phi(n)$ term in the formula for $\theta(n+1)$
in equations~\eqref{thetap}.

Numerical experimentation (specifically with the forced van der Pol oscillator 
considered in~\cite{GM04}) shows that the modified definition of
$p$ in~\eqref{modp} is inferior for long data sets with no noise when
only one choice of $c$ is used.   This was our reason for including
the extra $\phi(n)$ term in~\eqref{thetap}.
However, this term is no longer advantageous when $100$ choices of $c$
are sampled.  Moreover,~\eqref{modp} now has the advantage that 
$p(n)$ scales linearly with the data and so is far less susceptible to 
measurement noise than the test in~\cite{GM04}.
%\end{rmk}

\clearpage

%%%%%%%%%%%%%%%%%%%%%%%%%%%%%%%%%%%%%%%%%%%%%%%%%%%%%%%%%%%%%%%%%%%
%%%%%%%%%%%%%%%%%%%%%%%%%%%%%%%%%%%%%%%%%%%%%%%%%%%%%%%%%%%%%%%%%%%

\section{Traditional methods} 
\label{sec-tradmeth}

Suppose that
 $\phi(n)$ is the
data set originating from an unknown deterministic dynamical system.
(As usual $\phi$ is some one-dimensional observable of the system under
investigation.)
To compute the maximal Lyapunov exponent,
it is first necessary to reconstruct the dynamics~\cite{Takens81}. 
Define the $m$-dimensional {\it delay vector} 
\begin{align*}
{\bf \xi}_n=\{\phi(n),\phi(n+\tau),\phi(n+2\tau),\cdots,\phi(n+(m-1)\tau)\}.
\end{align*}
Here, the delay time $\tau>0$ is chosen to be an integer.  
If the underlying dynamics ${\bf x}_n$
lies inside a $d$-dimensional phase space, then
the delay reconstruction map ${\bf x}_n\mapsto\xi_n$ is an 
embedding provided $m>2d+1$ (corresponding to Whitney's embedding theorem
\cite{Whitney36}). 
%% An embedding is a smooth one-to-one map with a smooth
%% inverse. In particular, the metric properties in both spaces, the
%% $d$-dimensional $\{\bf \vec{x}\}$ and the ($2d+1$)-dimensional $\{\bf
%% \vec{\xi}\}$, are the same in the sense that distances in $\{\bf
%% \vec{x}\}$ and $\{\bf \vec{\xi}\}$ have a ratio which is uniformly
%% bounded, and bounded away from zero. This means that one can measure
%% the separation of trajectories also in the reconstructed phase space
%% in order to determine Lyapunov exponents. 
Sauer {\it et al.}~\cite{Sauer91,Sauer93} extended Takens' work by 
showing that it suffices to take $m>2d_0(A)$ where $d_0(A)$ is the fractal
box-counting dimension of the attractor.
Under these conditions, the reconstructed dynamics $\xi_1,\xi_2,\ldots$
in $\R^m$ faithfully represents the underlying dynamics.

The maximal Lyapunov exponent for a map $\xi_{n+1}=f(\xi_{n})$ is defined to be
\begin{align}
\label{lyap}
\lambda =  \lim_{n \to \infty} \frac1n
           \log(\|Df^n_{\xi_1}\|).
\end{align}
%% where $y_n$ is an infinitesimal displacement of an unperturbed orbit
%% in the direction of $y_0$ whose dynamics is given by
%% $y_{n+1}=Df(x_n)y_n$.  Hence, $y_n = Df^n(x_0)y_0$ where
%% $Df^n(x_0)=Df(x_{n-1})Df(x_{n-2})\cdots
%% Df(x_0)$.
%% 
There are (at least) two main numerical approaches for computing the 
maximal Lyapunov exponent: the {\it tangent space method} and 
the {\it direct method}.   

In tangent space methods, a model is fitted to the data to approximate
the Jacobian. By~\eqref{lyap}, the maximal Lyapunov exponent is defined as
the product of the Jacobian of the the linearised flow along the given 
trajectory.
If the underlying equations are not known the
Jacobian can be approximated. Such tangent space methods were first
introduced by Eckmann \& Ruelle \cite{EckmannRuelle86}, and then
further developed by Sano \& Sawada \cite{SanoSawada} and Eckmann {\it
et al.}~\cite{Eckmann86}.

Direct methods use the fact that nearby trajectories separate on
average asymptotically at rate $e^{\lambda n}$.
The original algorithm by Wolf~{\em et al.}~\cite{Wolf85} allows 
the computation of the whole spectrum of Lyapunov exponents. 
Rosenstein {\it et al.}
\cite{Rosenstein93} and Kantz \cite{Kantz94} concentrated
attention on the maximal Lyapunov exponent, and developed a more robust
method. 
In the method by Rosenstein {\it et al.}~\cite{Rosenstein93} for each
$m$-dimensional vector $\xi$ in the reconstructed phase space, its nearest 
neighbour $\xi^{\star}$ is determined.  Rosenstein {\it et al.} 
calculate $C(k)=<\log |f^k(\xi)-f^k(\xi^\star)|>$ where the angles
denote averaging over all $\xi=\xi_1,\xi_2,\ldots$ 
The function $C(k)$ shows roughly three
different regimes. An initial regime of flat increase, a subsequent
interval with exponential behaviour, and finally  a plateau
(because the separation cannot go beyond the extension of the
attractor). The maximal Lyapunov exponent is determined by the slope of
$C(k)$ in the usually quite short range of exponential 
behaviour.  

%\begin{rmk}
{\bf Remark 3.1} There are a number of inherent problems of both the
tangent space method and the direct method, linked to phase space
reconstruction. These difficulties, which include the choice of
embedding dimension $m$ and delay parameter $\tau$, are
well-documented and we refer to Casdagli {\it et
al.}~\cite{Casdagli91} and Schreiber \& Kantz \cite{SchreiberKantz95}
for detailed discussions.
%\end{rmk}

%%%%%%%%%%%%%%%%%%%%%%%%%%%%%%%%%%%%%%%%%%%%%%%%%%%%%%%%%%%%%%%%%%%
%%%%%%%%%%%%%%%%%%%%%%%%%%%%%%%%%%%%%%%%%%%%%%%%%%%%%%%%%%%%%%%%%%%

\section{Comparison with tangent space methods}
\label{sec-logistic}

In this section, we compare our test with the tangent space
method. Note that the definition of the maximal Lyapunov exponent
(\ref{lyap}) is itself a tangent space method. We use the method
developed by Sano \& Sawada \cite{SanoSawada}. In particular we study
the influence of measurement noise. 

As discussed in Section~\ref{sec-tradmeth},
tangent space methods seek to approximate the Jacobian $Df^n(x_0)$.
In the case of measurement (additive) noise,
the noise is amplified by higher order nonlinearities.
Similarly, if the dimensionality of the underlying dynamical
system is large, the evaluation of the diagonalised Jacobian
requires extensive multiplication of the
individual matrix elements of the Jacobian, and 
noise is again amplified by this process.

Neither of these problems arises in our test. Measurement noise
only enters our diagnostic variable $p(n)$ linearly via the
observation $\phi(n)$ irrespective of the underlying dynamical
system. 

We consider an illustrative example.
Our contaminated data has the form
$\tilde\phi(n)=\phi(n) + ({\cal N}/100)\eta_n$ where $x_n$ is the clean data,
${\cal{N}}$ is the
noise-level in percent, and $\eta_1,\eta_2,\ldots$ are i.i.d.\
random variables drawn from a uniform
distribution on $[-1,1]$. We note that the results are similar in the
case of normally distributed noise.  

As in Section~\ref{sec-test}, we study the logistic map
\begin{align}
x_{n+1}=\mu x_n(1-x_n),
\label{logi}
\end{align}
varying the parameter $\mu$ in the range $3.5\le\mu\le 4$
in increments of $0.001$. 
%% where a generic period-doubling route into chaos is
%% observed and several periodic windows appear. 
We again use data sets consisting of
$N=1,000$ iterates, after a transient of $20,000$ iterates starting
from the initial condition
$0.0001$.    Throughout, we take $\phi(n)=x_n$
as the observable. 

\begin{figure}[htb]
   \centerline{%
    \includegraphics[angle=-90,width=.5\textwidth]{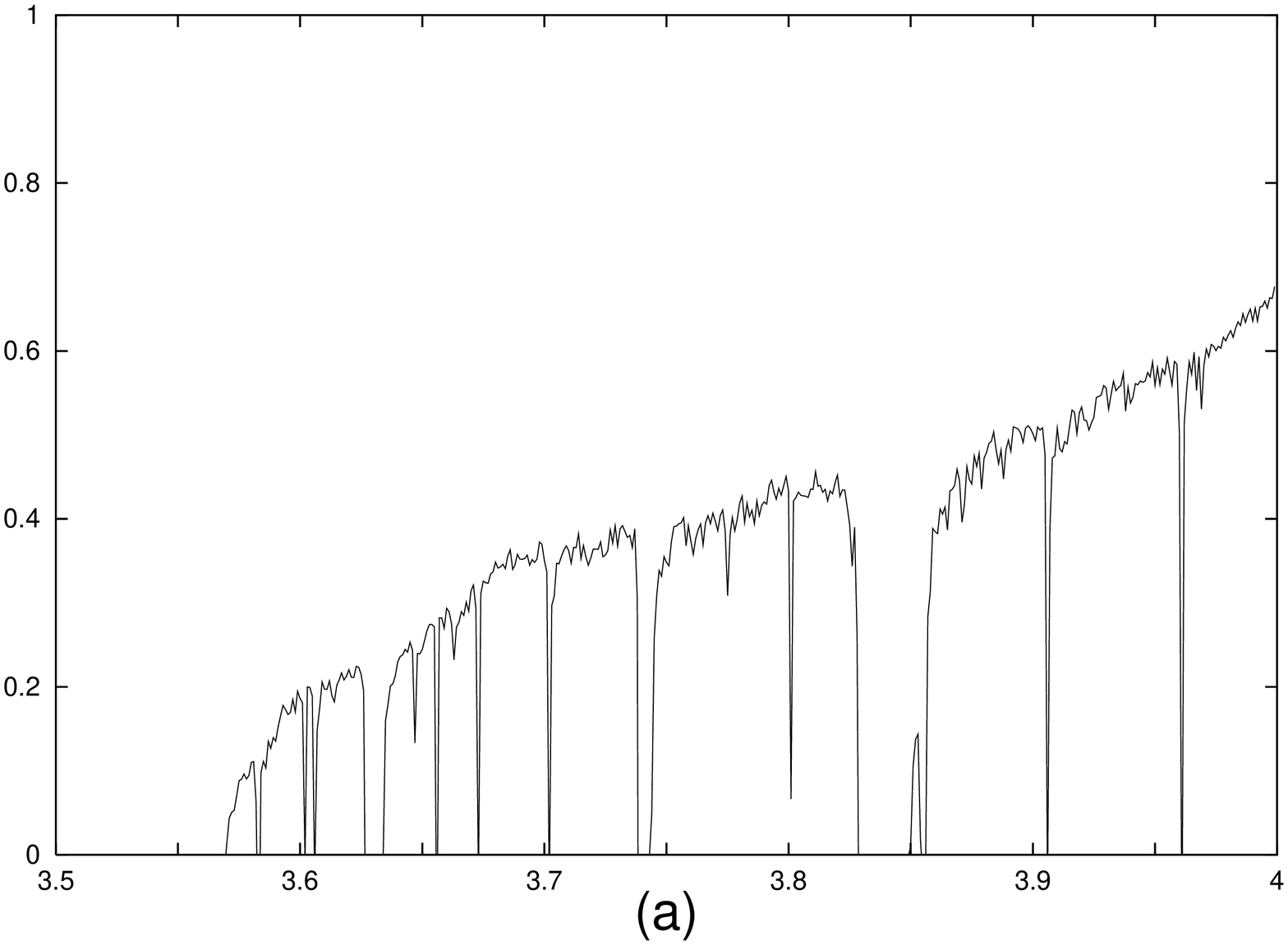}
    \includegraphics[angle=-90,width=.5\textwidth]{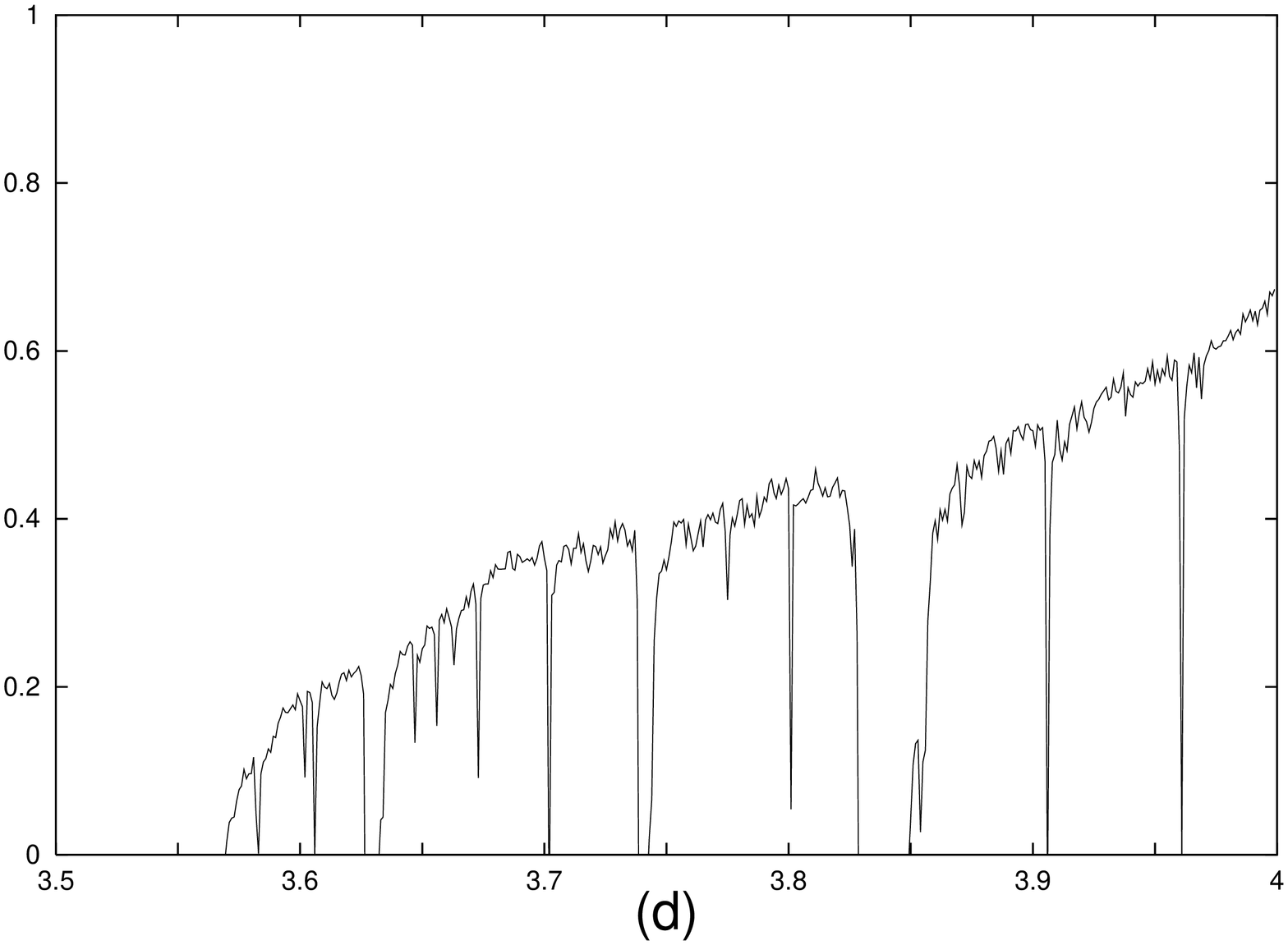}}
   \centerline{%
    \includegraphics[angle=-90,width=.5\textwidth]{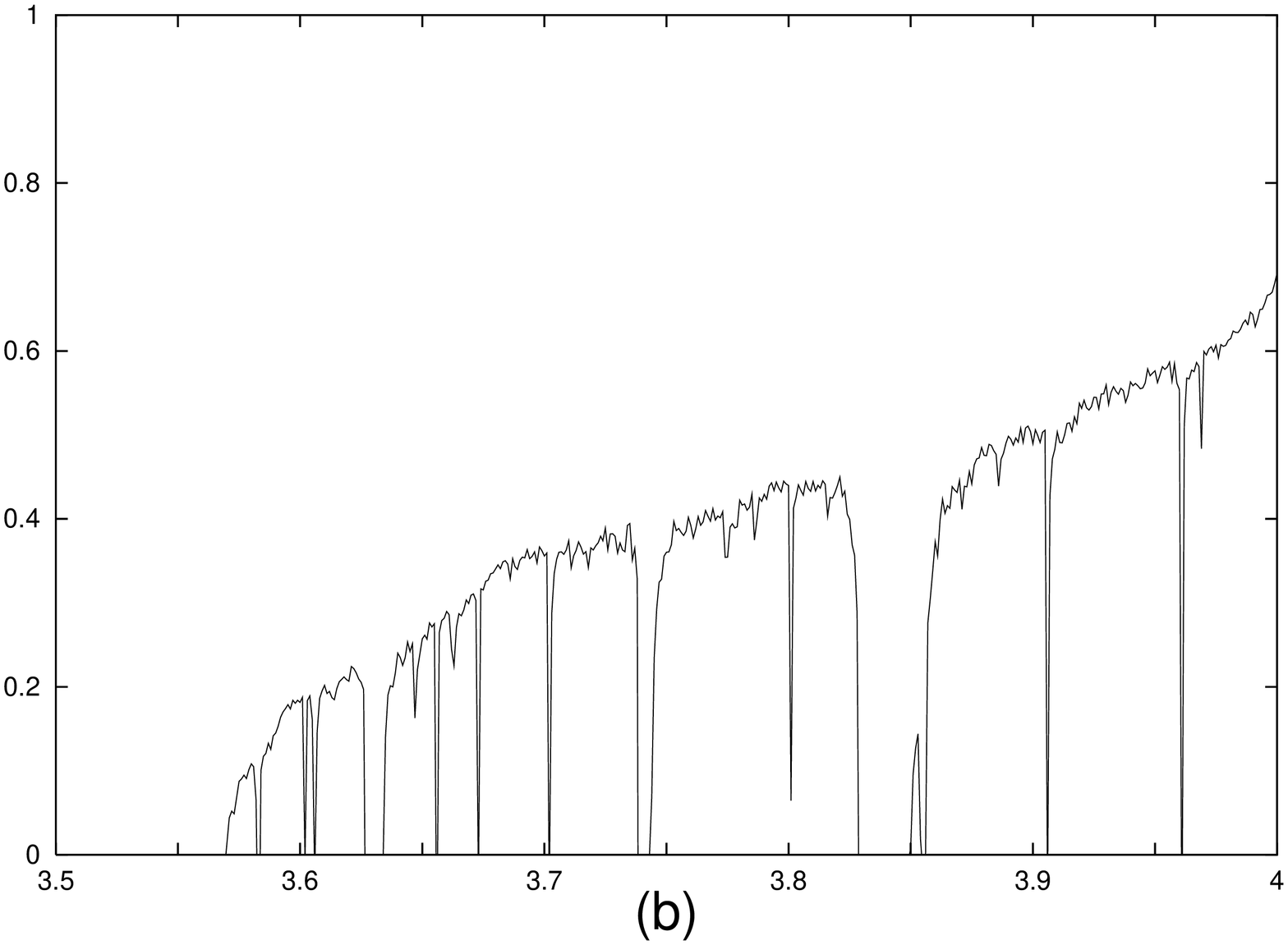}
    \includegraphics[angle=-90,width=.5\textwidth]{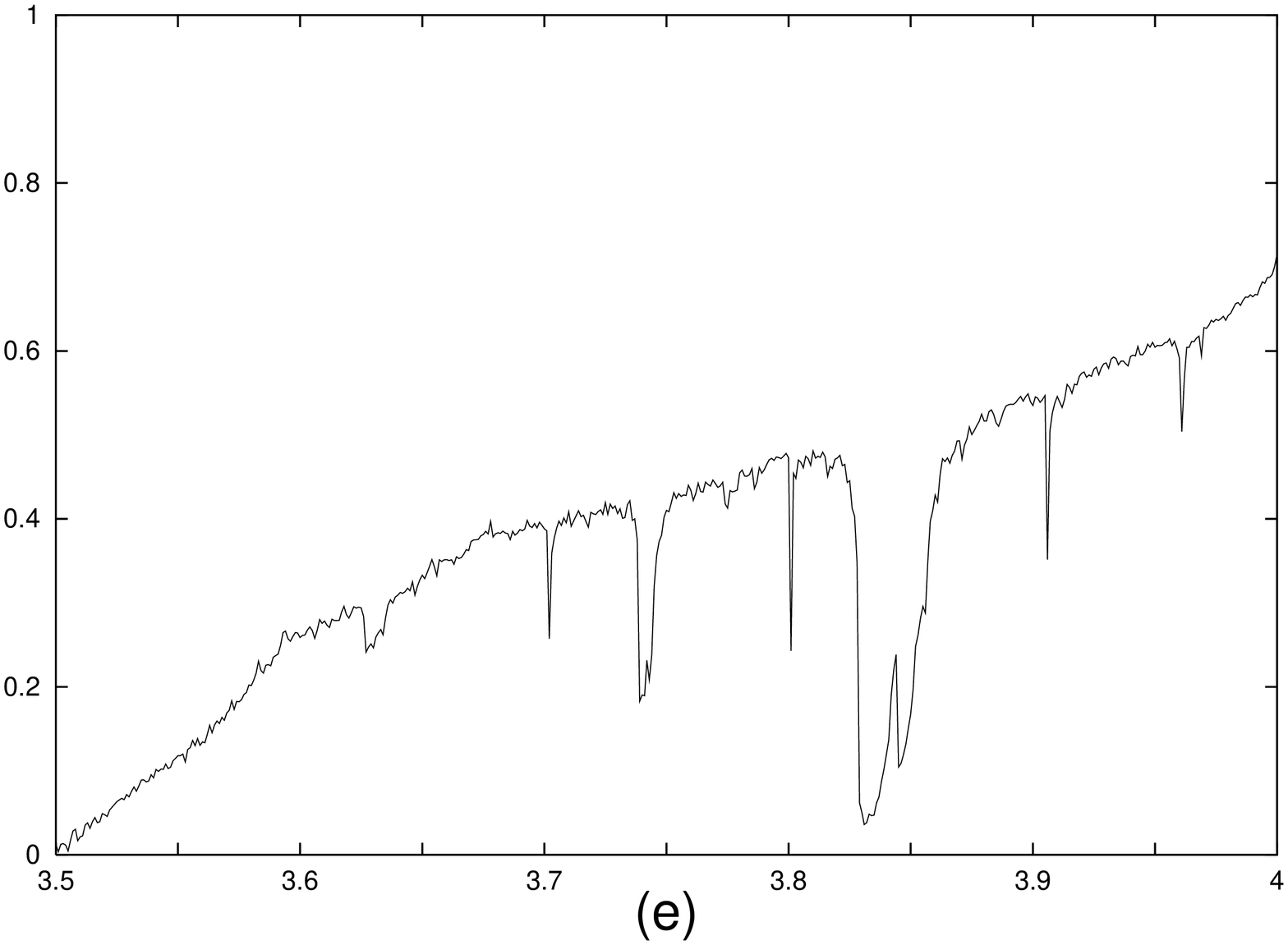}}
   \centerline{%
    \includegraphics[angle=-90,width=.5\textwidth]{FIGURES/K100.ps}
    \includegraphics[angle=-90,width=.5\textwidth]{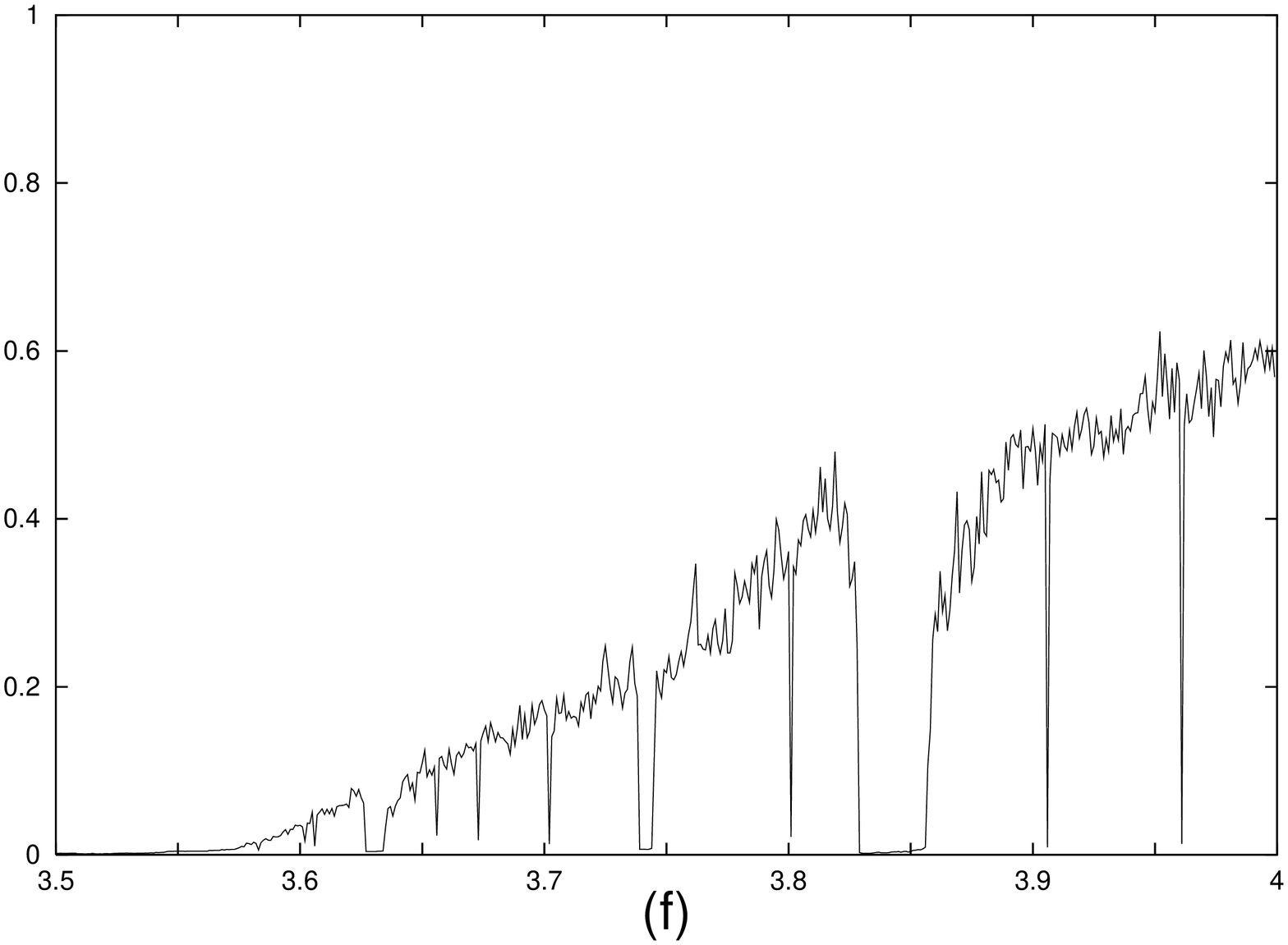}}
    \caption{Plots of the Lyapunov exponent and $K$
    versus $\mu$ for the logistic map~\eqref{logi} with 
    $3.5\le\mu\le 4$ using $1,000$ iterates. First row: ``Exact''
    Lyapunov exponent. Second row: Method by Sano \& Sawada with $m=2$ and
    $0.000001$ as an allowed distance to define nearest
    neighbours.  Third row:  Our test with $100$ different
    values of $c$.   The first column uses noise-free data.   
The second column incorporates 1\% measurement noise.}
 \label{fig-logi-combi} %\hrule
  \end{figure}

Our results for noise-free data are shown in the first column of
Figure~\ref{fig-logi-combi}.
As a benchmark, we compute the ``exact'' Lyapunov exponent
making use of the explicitly given form of the map in~\eqref{logi}, see
Figure~\ref{fig-logi-combi}(a), 
Next, we use the tangent space method proposed by Sano \& Sawada
\cite{SanoSawada} 
for approximating
the dynamics on the tangent space using phase-space reconstruction, see
Figure~\ref{fig-logi-combi}(b). 
The results from our modified test (Section~\ref{sec-test})
 with $100$ values of $c$ are shown in Figure~\ref{fig-logi-combi}(c). 

%% It is clearly seen that the ``diagnostic'' methods,
%% i.e. the method by Sano \& Sawada \cite{SanoSawada} and our method,
%% perform equally well.

Now we add measurement noise with
noise-level ${\cal{N}}=1\%$. 
As can be seen from Figure~\ref{fig-logi-combi}(e), 
the phase space reconstruction
method by Sano \& Sawada \cite{SanoSawada} performs rather
poorly. Moreover, the numerical values and the overall qualitative
behaviour of the maximal Lyapunov exponent with respect to varying
$\mu$ depends very sensitively on the particular choice of the
embedding dimension $m$ and the value chosen to define nearest
neighbours. 
In contrast, our test performs as  well as it did in
the noise-free case, see Figure~\ref{fig-logi-combi}(f).

\begin{figure}[htb]
   \centerline{%
    \includegraphics[angle=-90,width=.5\textwidth]{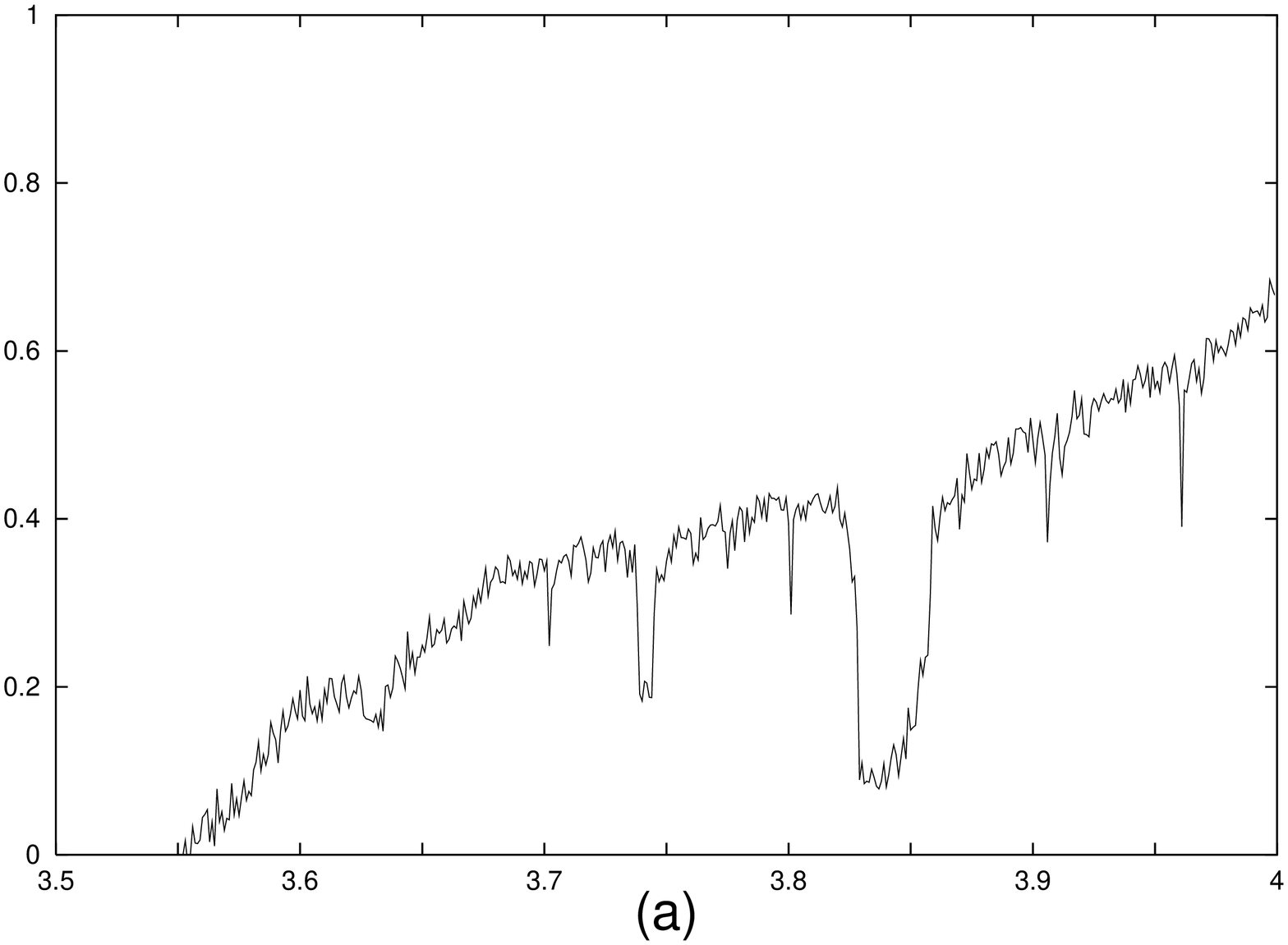}
    \includegraphics[angle=-90,width=.5\textwidth]{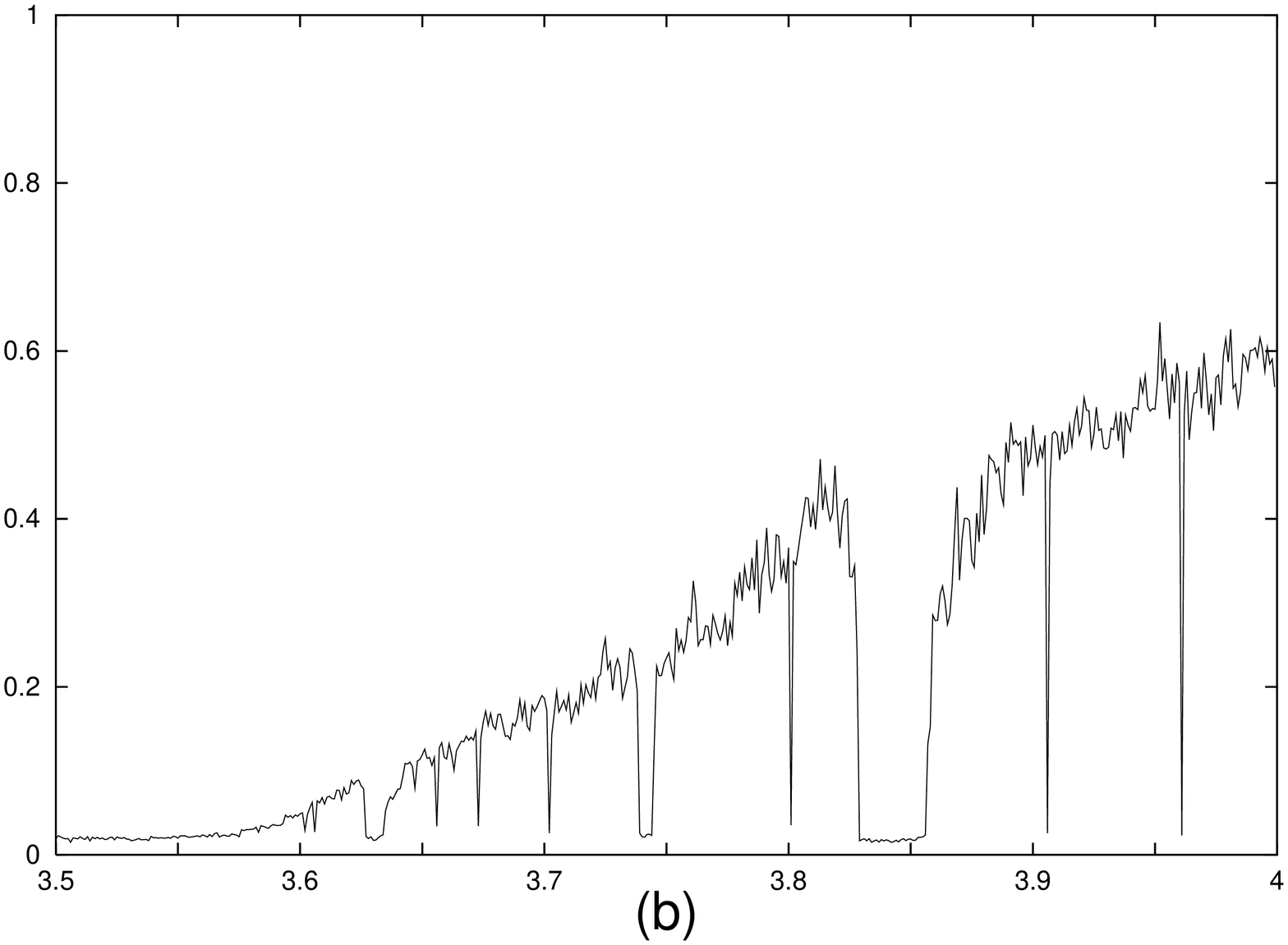}}
    \caption{Plots of the Lyapunov exponent and $K$
    versus $\mu$ for the logistic map~\eqref{logi} with 
    $3.5\le\mu\le 4$ using $1,000$ iterates with $10\%$ noise. 
(a): ``Exact'' Lyapunov exponent. 
    (b): Our test with $100$ different
    values of $c$.  } 
 \label{fig-logi10} %\hrule
  \end{figure}

It turns out that the tangent space method copes poorly with measurement
noise even when phase space reconstruction is not required.
To see that this is the case, we fed contaminated data into the
exact expression for the Jacobian computed analytically from~\eqref{logi}.
In Figure~\ref{fig-logi10}, we show the results for the ``exact''
tangent space  method
compared with our test, both with $10\%$ measurement noise.
This seems to be  conclusive evidence that our test is better than the 
tangent space method.    We have also verified that our test deals
comfortably with $20\%$ noise.

There is another advantage of our method, even in the noise-free
case. Suppose that a whole scan through the bifurcation parameter is
not available but instead data is only available for one particular
value of the bifurcation parameter. A value of $0.2$ for the maximal
Lyapunov exponent is inconclusive (see for example the periodic windows in
Figure~\ref{fig-logi-combi}), whereas a value of $K=0.01$ indicates
regular dynamics.  In this sense, our test is
a better absolute test than the Lyapunov exponent.

\clearpage

%%%%%%%%%%%%%%%%%%%%%%%%%%%%%%%%%%%%%%%%%%%%%%%%%%%%%%%%%%%%%%%%%%%%
%%%%%%%%%%%%%%%%%%%%%%%%%%%%%%%%%%%%%%%%%%%%%%%%%%%%%%%%%%%%%%%%%%

\section{Comparison with direct methods}
\label{sec-rosenstein}

In this section, we compare our test for chaos against the direct method
for computing the maximal Lyapunov exponent.  In particular, we use the
direct method proposed by Rosenstein {\it et al.}~\cite{Rosenstein93}. 

The standard Lorenz equation is a $3$-dimensional truncation of a
certain system of partial differential equations (PDEs). This is too
simple a system to present a challenge for testing for chaos. 
Hence, we study the $n$-dimensional Lorenz system, known as the Lorenz~$96$
system:
\begin{align}
\label{lorenz96}
\frac{dx_i}{dt}=x_{i-1}(x_{i+1}-x_{i-2})-x_i+r \quad {\rm{with}} \quad
i=1,\cdots,n\; .
\end{align}
This system of ODEs was first introduced by Lorenz as an idealised 
model for the atmosphere \cite{Lorenz96}, and has been studied by
Orrell \& Smith \cite{OrrellSmith03}. In the following we choose
$n=8$. For $3.8\le r \le 3.9$
there are periodic windows, whereas for $5.4\le r \le 6.8$ most regular
windows are due to quasiperiodic dynamics \cite{OrrellSmith03}.
Throughout,
we integrate the system using a time step of $0.05$ and calculate
until $250,000$ units of time
after a transient of $75,000$ units of time.
%75,00 = 30,000*nt(=50)*d
However, we use only the data recorded after each $2.5$ units of time,
yielding a time series of $N=10,000$ data points.
In this way, we obtain a discrete
time series mimicking the situation for experimental data. As an
observable we take $\phi=x_2+x_3+x_4$,
so $\phi(n)=x_2(t)+x_3(t)+x_4(t)$ with $t=2.5n$.

\begin{figure}[htb]
   \begin{center}
%lorenzN/Lpap0.expon
    \includegraphics[angle=-90,width=0.5\textwidth]{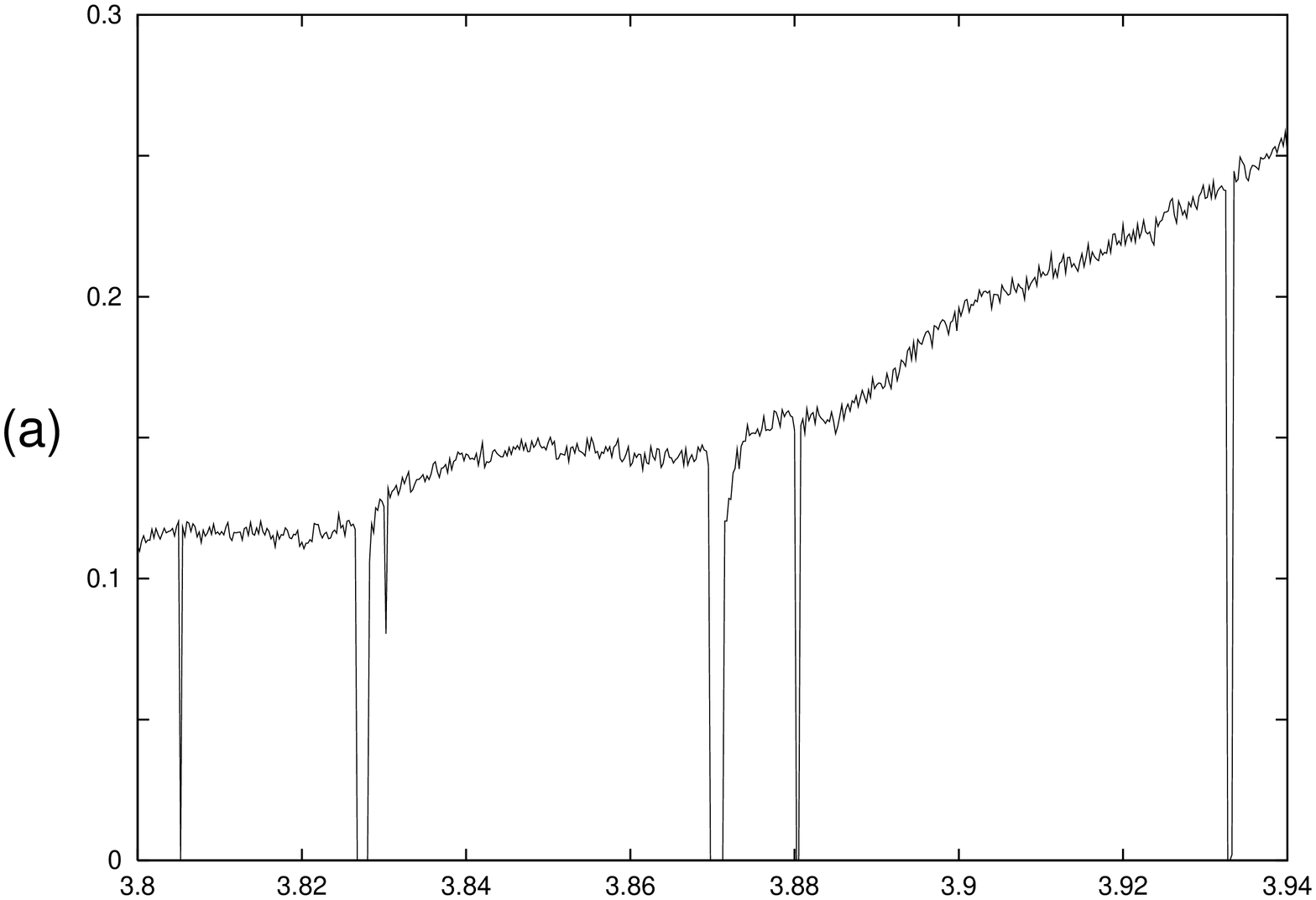}
%lorenzN/Reconstruction/Rpap8_0.expon
    \includegraphics[angle=-90,width=0.5\textwidth]{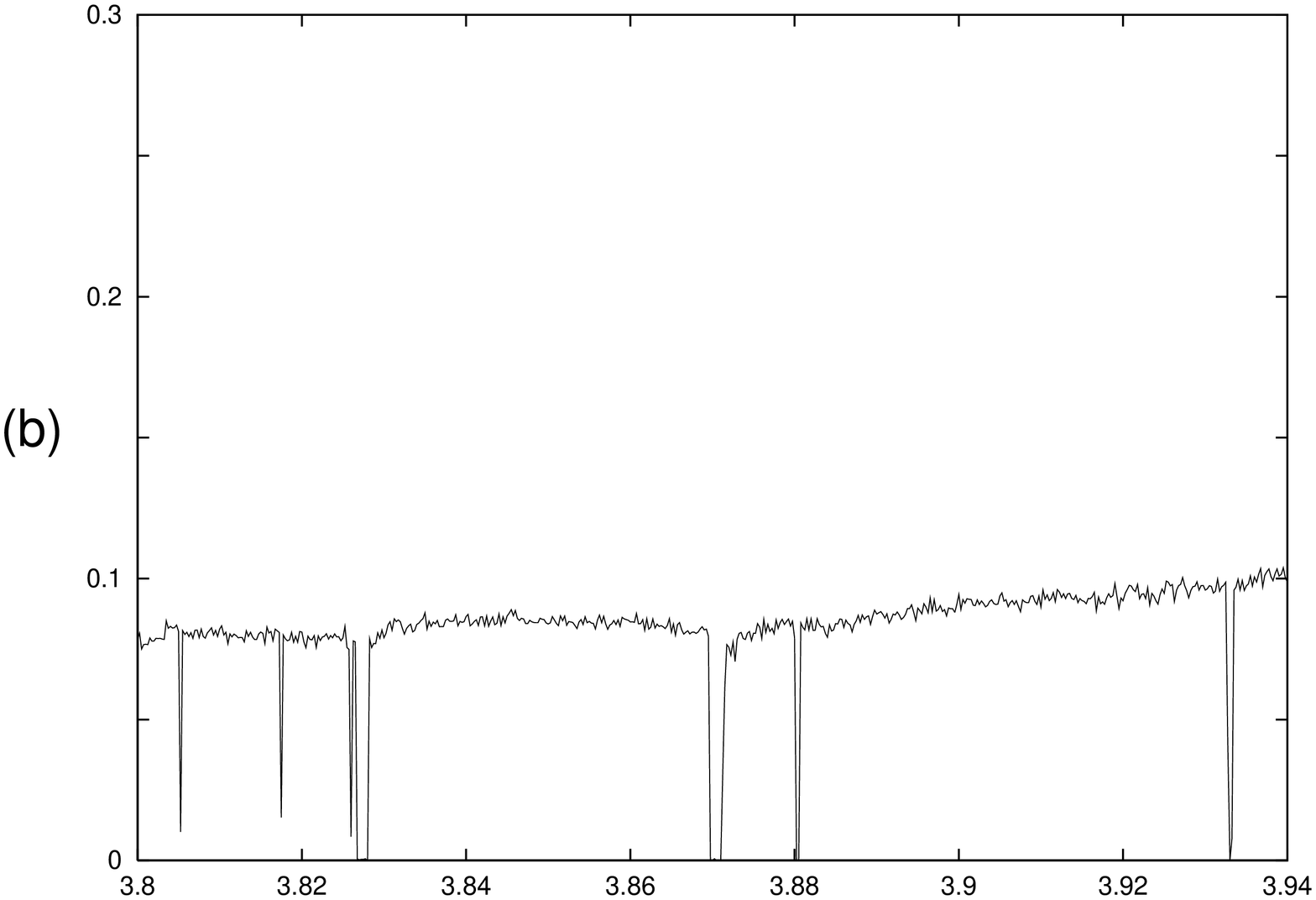}
%lorenzN/E2medianc_disc/Kpap8_0.slope
    \includegraphics[angle=-90,width=0.5\textwidth]{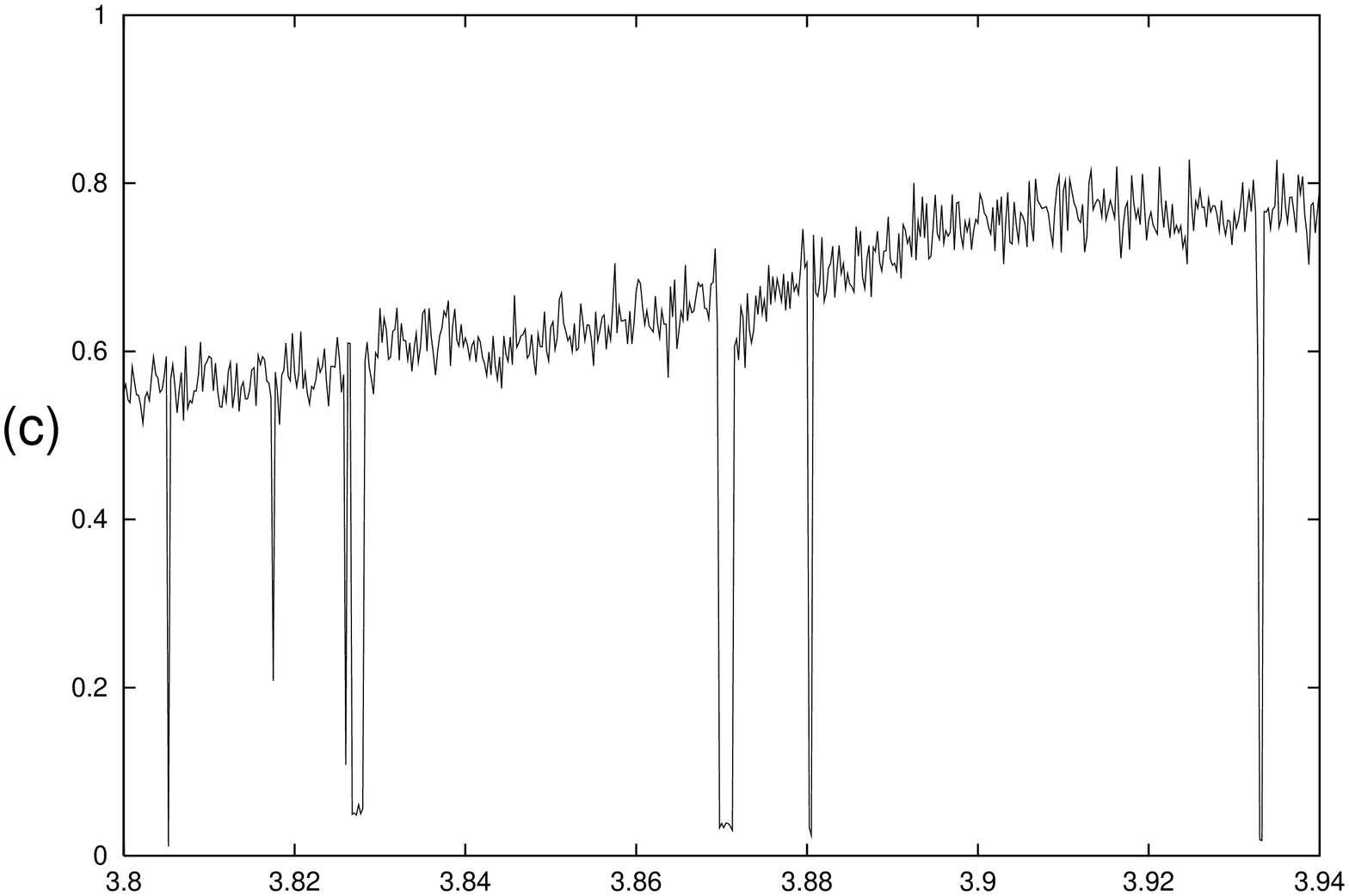}
    \end{center} 
    \caption{Plots of the maximal Lyapunov exponent and $K$ versus $r$ for the 
    $8$-dimensional Lorenz~$96$ system (\ref{lorenz96})
in the regime $3.8\le r\le 3.94$ using $N=10,000$ noise-free
data points. 
    (a): ``Exact'' maximal Lyapunov exponent.
    (b): Maximal Lyapunov exponent, direct method~\cite{Rosenstein93}.
    (c): Our test.}
 \label{fig-lorenz8}
 \end{figure}

We focus first on the range $3.8\le r\le 3.94$ with increments 
of $0.00025$. 
The ``exact'' maximal Lyapunov exponent calculated by solving the
analytic linearised flow for a noise free 
trajectory is shown in Figure~\ref{fig-lorenz8}(a) and serves again
as a benchmark to compare maximal Lyapunov exponents with our test.
%%Note that the maximal Lyapunov exponent for a flow is always nonnegative.

In Figure~\ref{fig-lorenz8}(b,c) we show the results for noise-free
data for the direct method and for our method.
The methods appear to work equally well here. It is
interesting to note that at $r=3.8175$ both methods
indicate regular behaviour whereas the ``exact''
maximal Lyapunov exponent is positive at that value of $r$ indicating
chaotic behaviour. As a matter of fact there is a periodic window close
by at  $r=3.817525$ which we checked by using the ``exact'' maximal
Lyapunov exponent. 

In Figure~\ref{fig-lorenz8}(b), we have taken the same vertical range
(from $0$ to $0.3$) as was natural in Figure~\ref{fig-lorenz8}(a).   
Note the poor convergence of the maximal Lyapunov exponent 
for the direct method.

\begin{figure}[htb]
   \begin{center}
%lorenzN/Reconstruction/Rpap8_10.expon
    \includegraphics[angle=-90,width=0.5\textwidth]{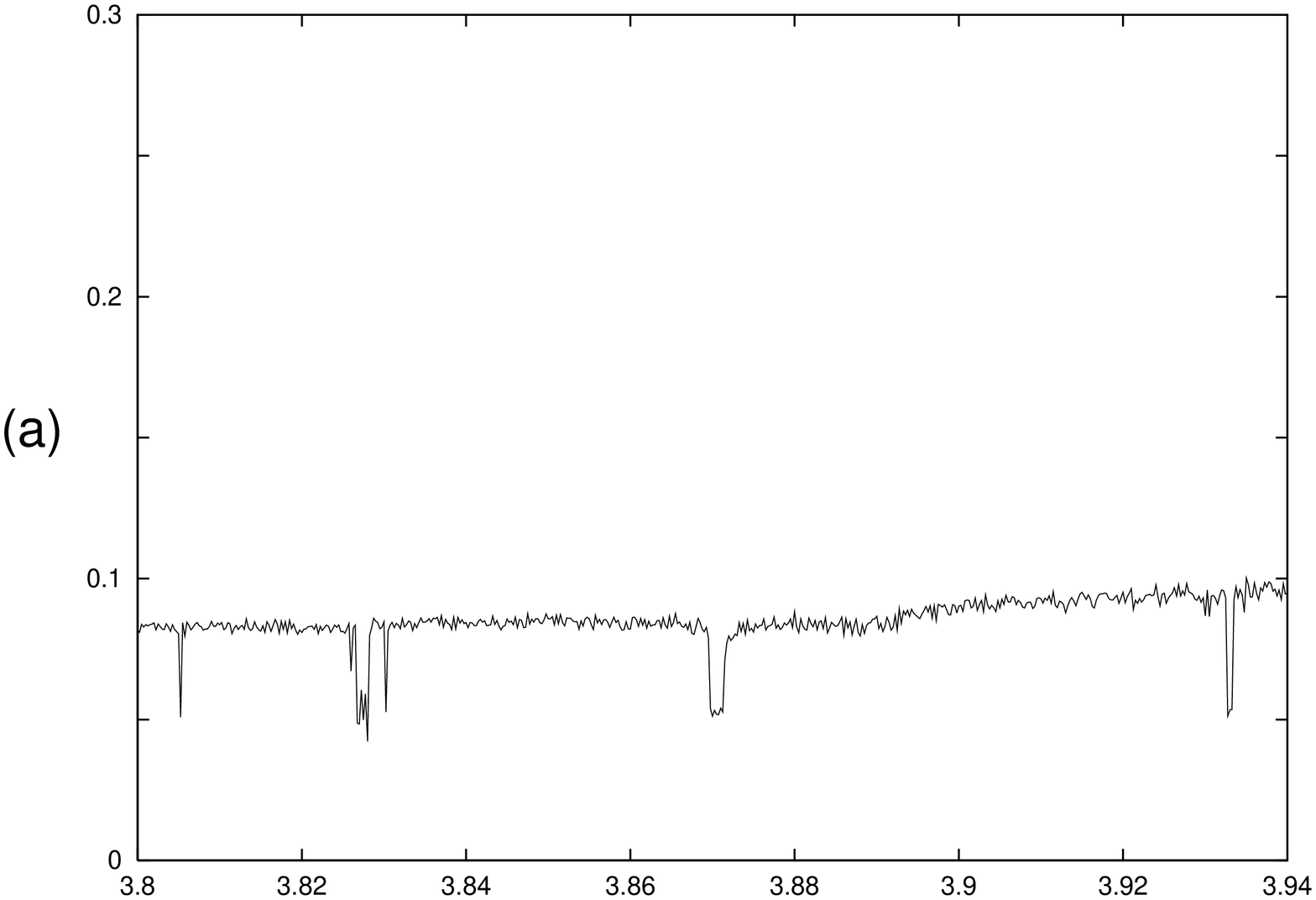}
%lorenzN/E2medianc_disc/Kpap8_10.slope
    \includegraphics[angle=-90,width=0.5\textwidth]{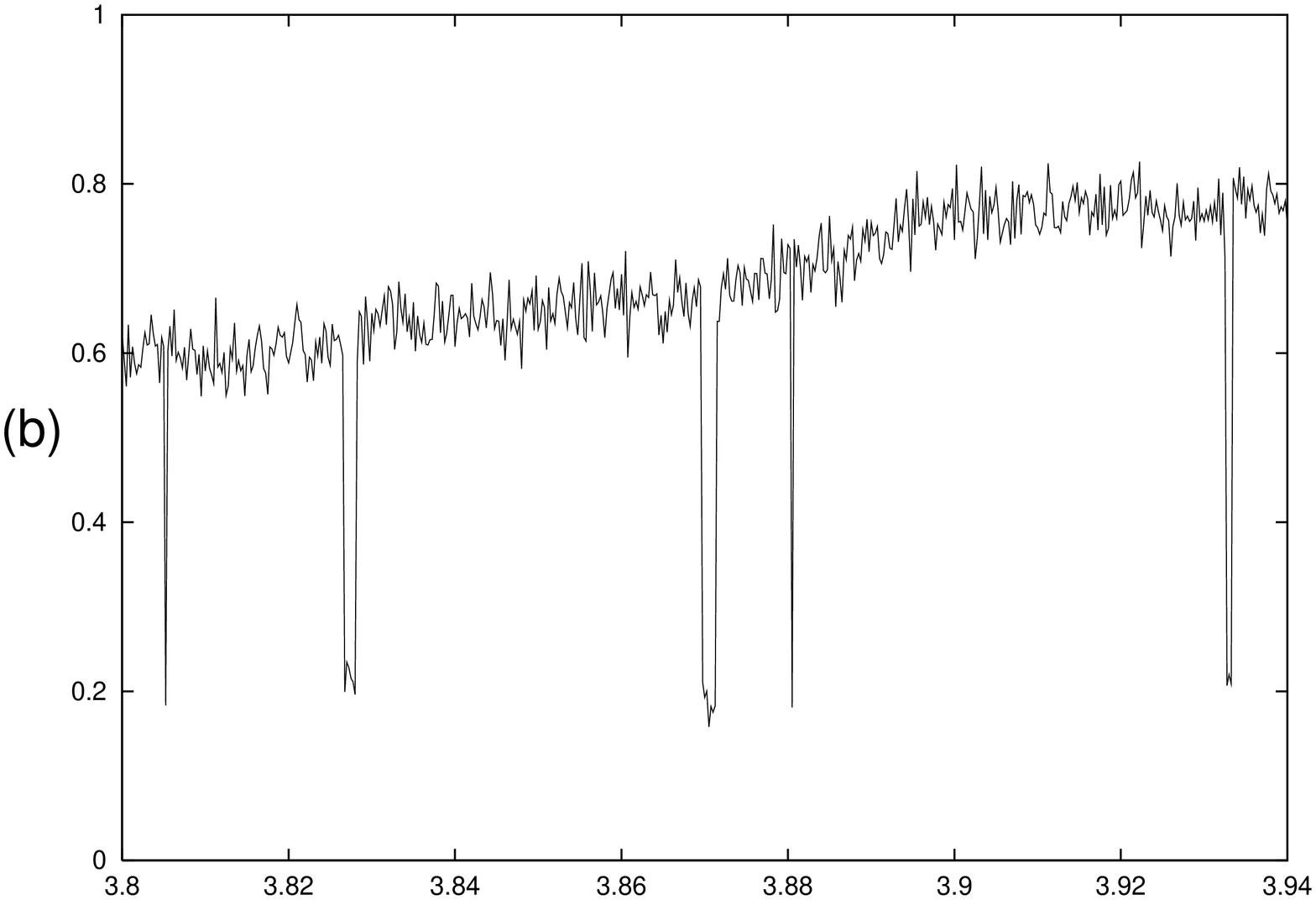}
    \end{center} 
    \caption{Plots of the maximal Lyapunov exponent and $K$ versus $r$ for the 
    $8$-dimensional Lorenz~$96$ system (\ref{lorenz96}) 
in the regime $3.8\le r\le 3.94$ using
$N=10,000$ data points with $10\%$ measurement noise. 
    (a): Maximal Lyapunov, direct method~\cite{Rosenstein93}. (b): Our test.}
 \label{fig-lorenz8-noise}
 \end{figure}

The results for i.i.d.\ measurement noise using a noise-level of $10\%$ are
shown in Figure~\ref{fig-lorenz8-noise}. The methods seem to perform 
roughly on an equal level. Whereas our method misses the periodic
window at $r=3.83025$, the direct method misses the periodic window at
$r=3.88025$.   

\clearpage

\begin{figure}[htb]
   \begin{center}
%lorenzN/LpapQ0.expon
    \includegraphics[angle=-90,width=0.5\textwidth]{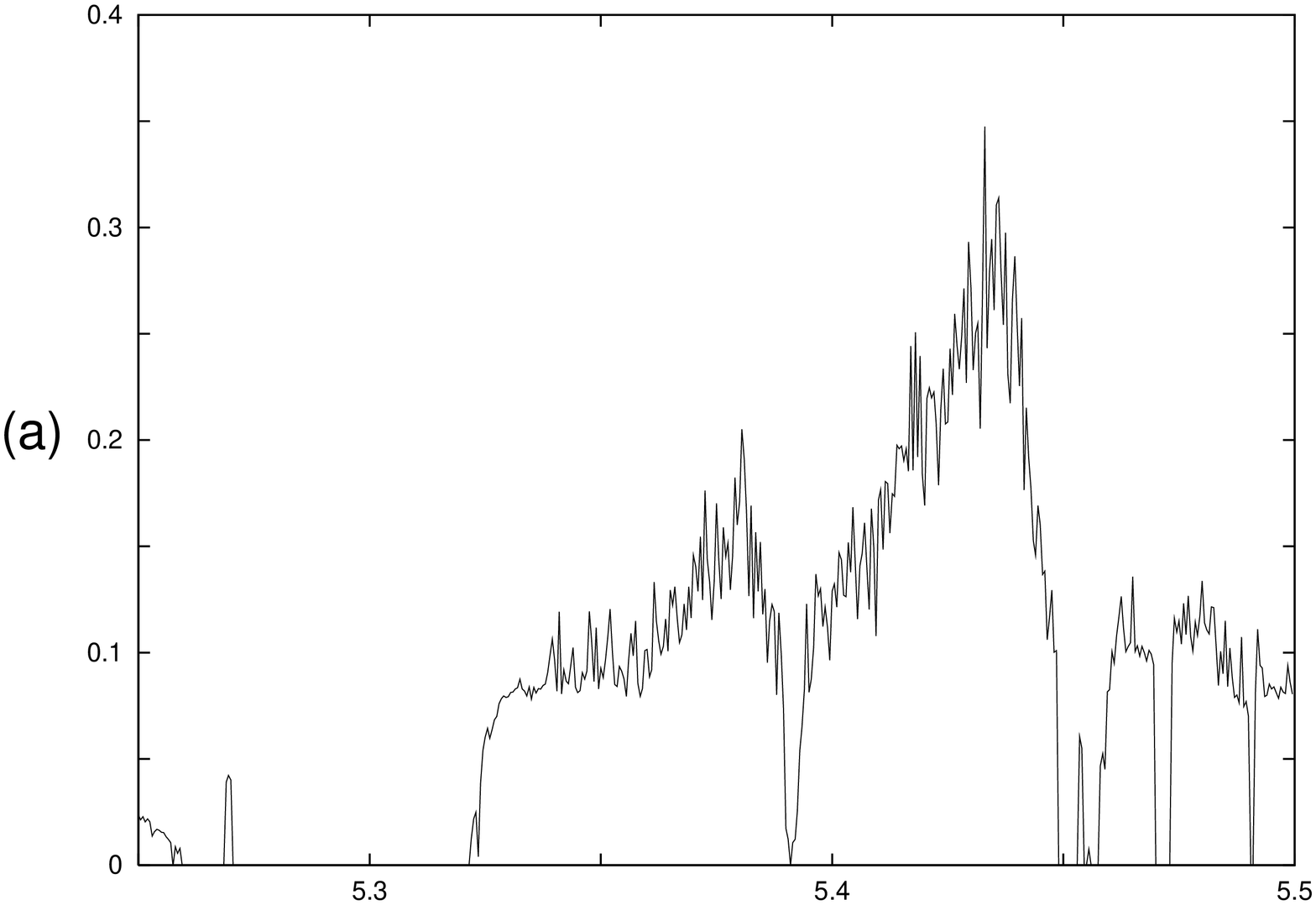}
%lorenzN/Reconstruction/RpapQ_0.expon
    \includegraphics[angle=-90,width=0.5\textwidth]{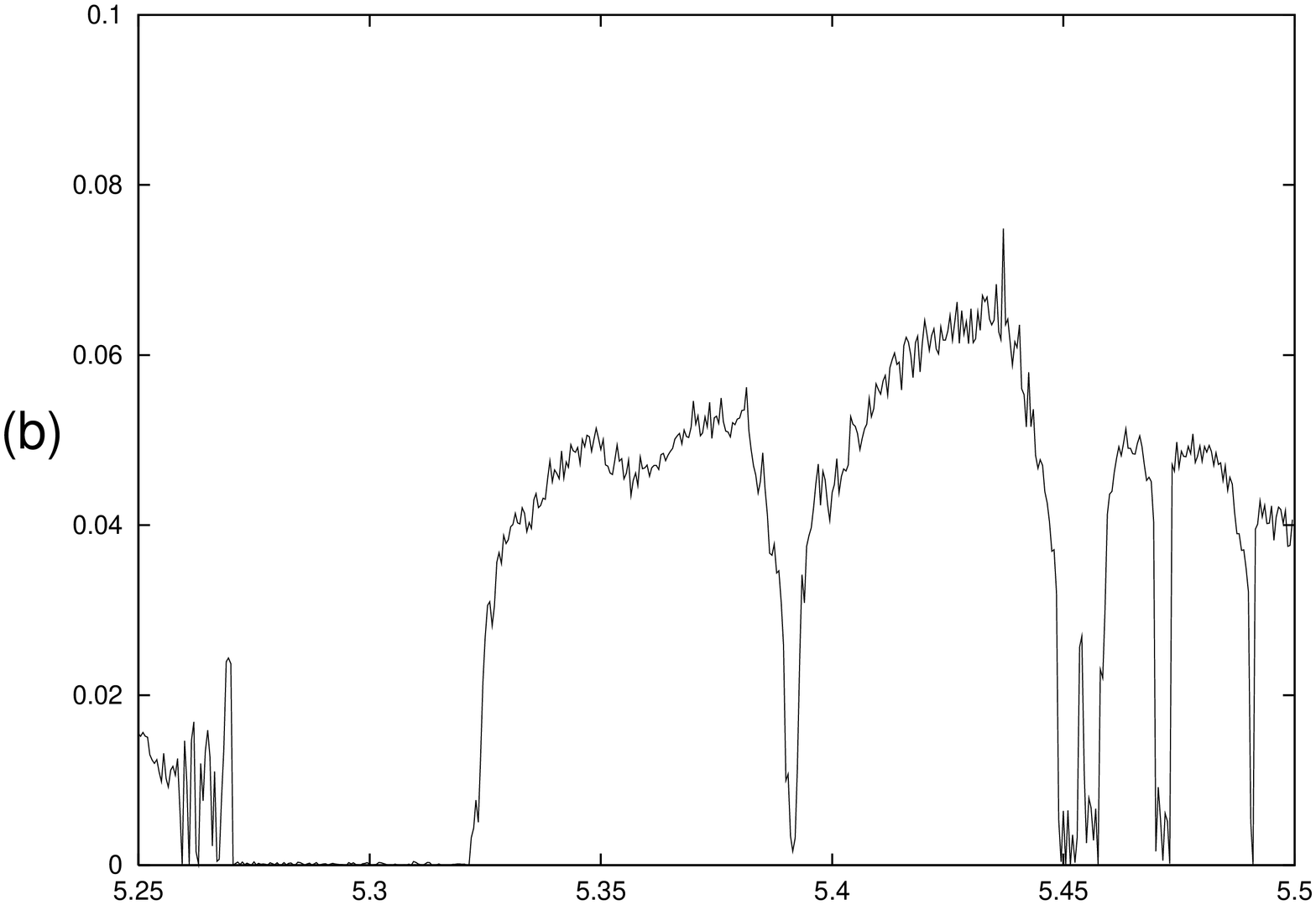}
%lorenzN/E2medianc_disc/KpapQ_0.slope
    \includegraphics[angle=-90,width=0.5\textwidth]{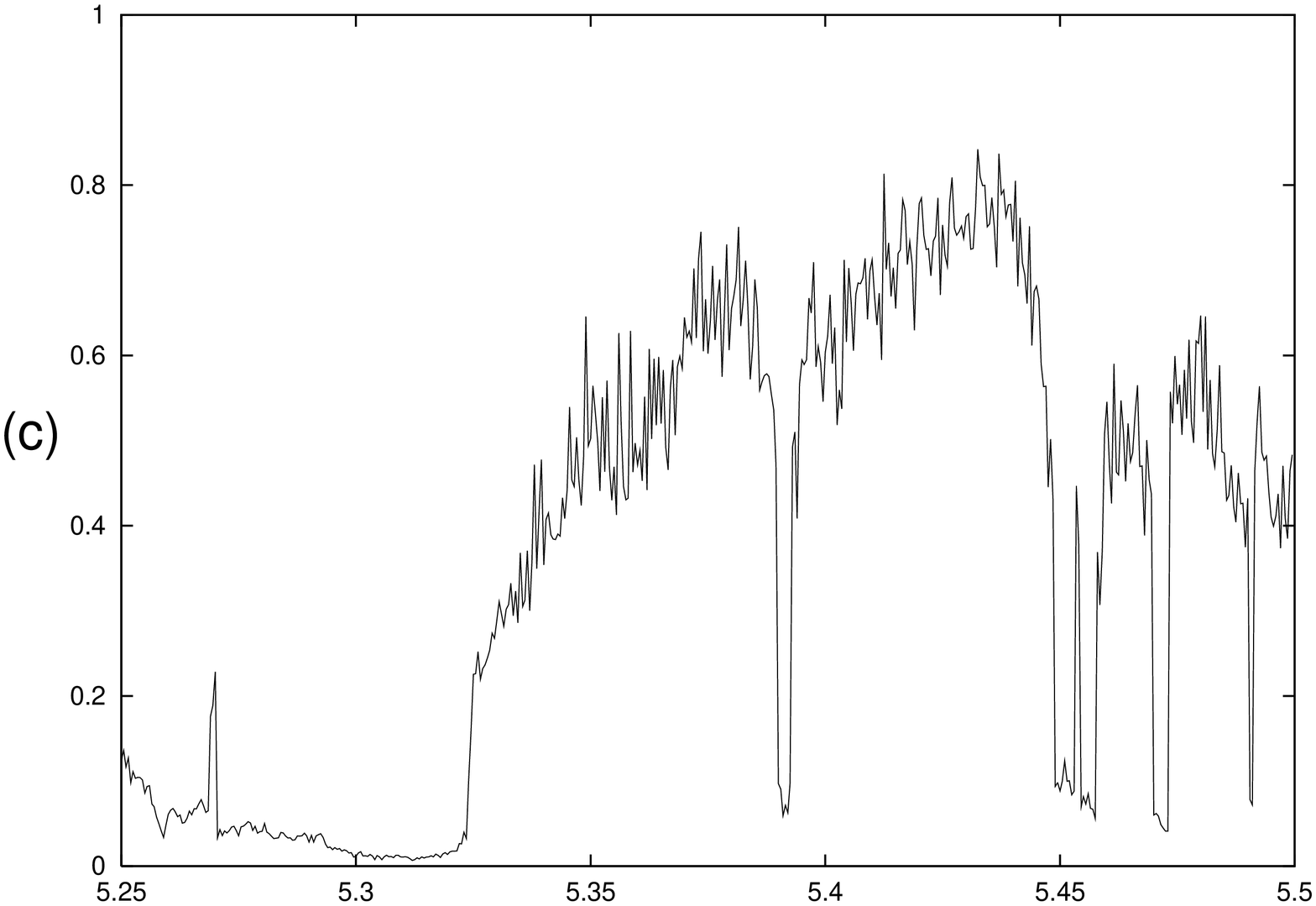}
    \end{center} 
    \caption{Plots of the maximal Lyapunov exponent and $K$ versus $r$ for the 
    $8$-dimensional Lorenz~$96$ system (\ref{lorenz96}) in
    the quasiperiodic regime $5.25\le r\le 5.5$ 
    using $N=10,000$ noise-free data points.
    (a): ``Exact'' maximal Lyapunov exponent.
    (b): Maximal Lyapunov exponent, direct method~\cite{Rosenstein93}. 
    (c): Our test.}
 \label{fig-lorenz8-QP}
 \end{figure}

Next, we consider the  quasiperiodic
regime $5.25\le r\le 5.5$ increasing in increments of $0.0005$.
The ``exact'' maximal Lyapunov exponent is shown in
Figure~\ref{fig-lorenz8-QP}(a).
In Figure~\ref{fig-lorenz8-QP}(b,c), we show the results for noise-free
data for the direct method and for our method.
Note the difference in the vertical
ranges in Figure~\ref{fig-lorenz8-QP}(a,b).
Whereas the $K$ test shows a distinction of regular quasiperiodic
motion and chaotic motion near $r=5.25$ the direct method cannot
properly resolve this. This becomes more drastic in the presence of noise.

 \begin{figure}[htb]
   \begin{center}
%lorenzN/Reconstruction/RpapQ_10.expon
    \includegraphics[angle=-90,width=0.5\textwidth]{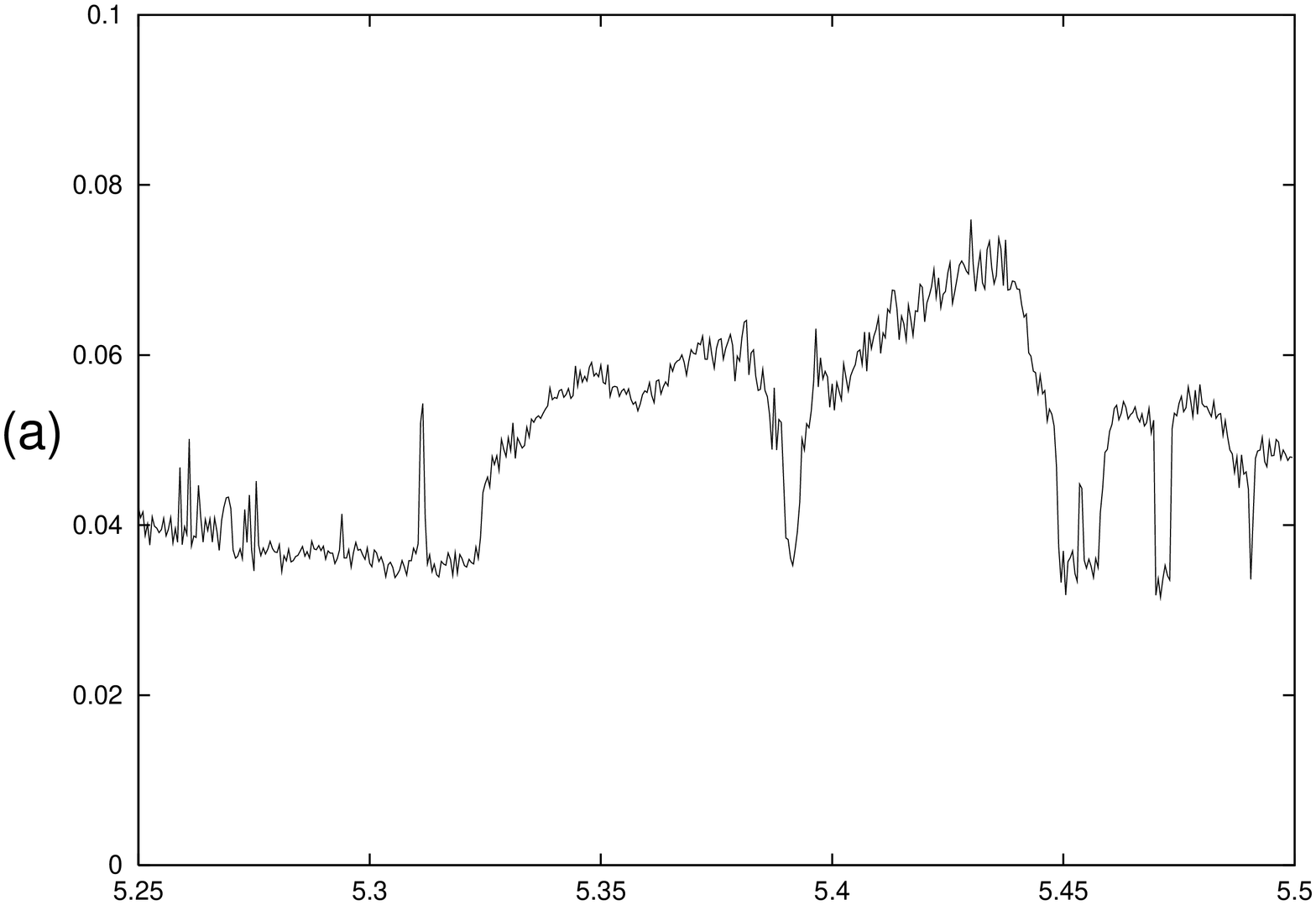}
%lorenzN/E2medianc_disc/KpapQ_10.slope
    \includegraphics[angle=-90,width=0.5\textwidth]{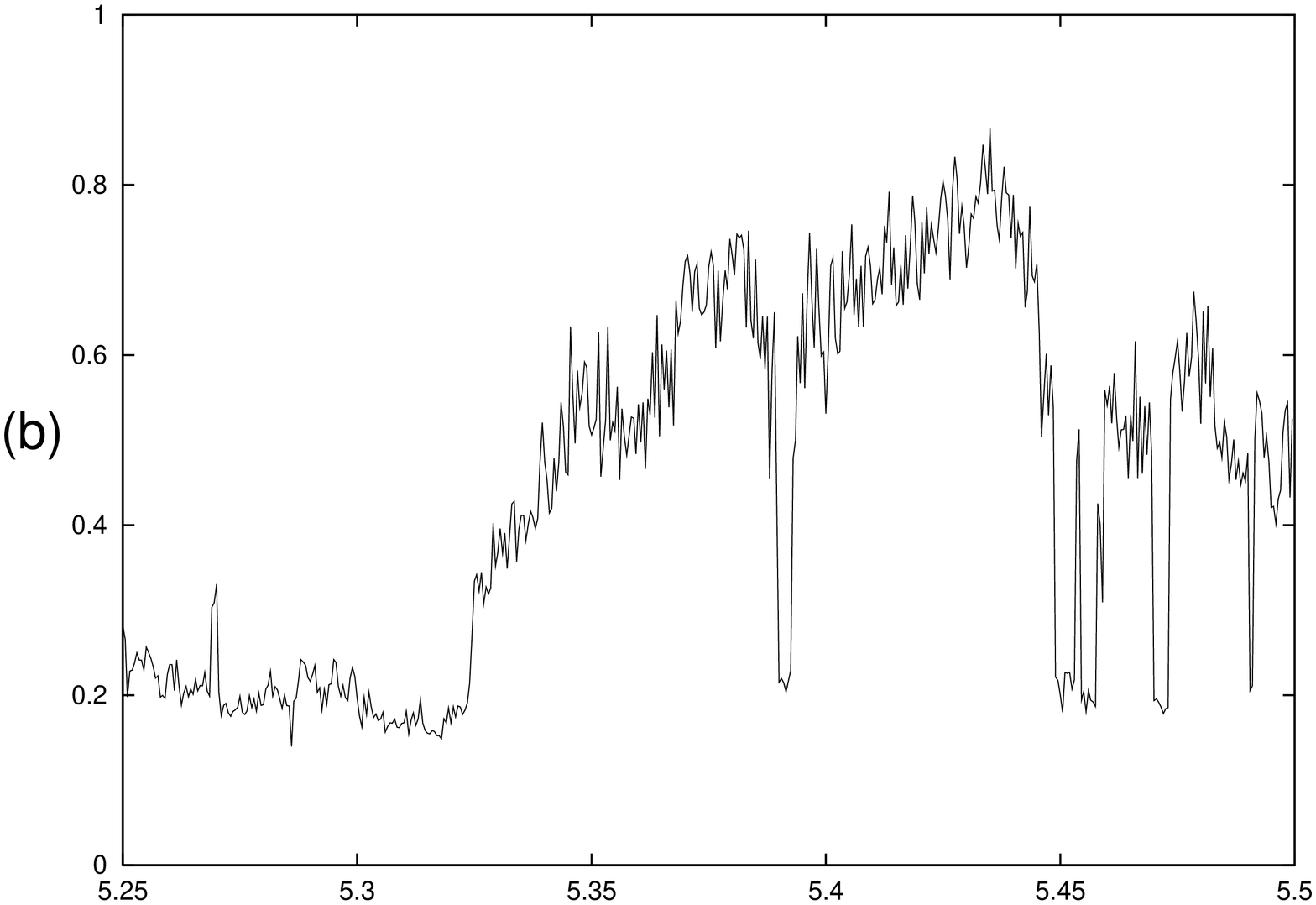}
    \end{center} 
    \caption{Plots of the maximal Lyapunov exponent and $K$ versus $r$ for the 
    $8$-dimensional Lorenz~$96$ system (\ref{lorenz96}) 
    added measurement noise in the quasiperiodic regime $5.25\le r\le 5.5$
    using $N=10,000$ data points with $10\%$ measurement noise.
    (a): Maximal Lyapunov exponent, direct method~\cite{Rosenstein93}. 
    (b): Our test.
    }
 \label{fig-lorenz8-QP-noise}
 \end{figure}

In Figure~\ref{fig-lorenz8-QP-noise}, we show the results
for the quasiperiodic regime when $10\%$ measurement noise is added.
Here the distinction near $r=5.25$ is not possible at all for the
direct method. The $K$ test shows a faint signature close to $r=5.25$
but the chaotic interval $5.269\le r\le 5.27$ is clearly visible. Moreover,
the direct method has a spurious chaotic interval $5.311\le r \le 5.312$.  
In addition, the direct method fails as an absolute test. For
example the periodic region neighbouring the chaotic peak at
$r=5.4535$ is almost indistinguishable from an absolute point of view.
The Lyapunov exponent of the periodic regime is about $80\%$ the value
of the chaotic peak. In contrast our method does not have any spurious
regular or chaotic intervals. Also it works well as
an absolute test and the difference between values of $K$ indicating regular
dynamics and values of $K$ indicating chaotic dynamics are well separated.

\clearpage

\section{Summary}
We have presented a modification of our test for chaos \cite{GM04}
which better handles moderate amount of contaminated data.
In particular we have simplified the diagnostic equation,
and we now compute the asymptotic growth
rate $K$ as the median value over realizations over many different
values of $c$. 

We have shown that our test is much better at
coping with measurement noise than tangent space methods.
%% The higher the order of nonlinearity, the greater the advantage
%% of the $K$ test. The explanation for this is quite straightforward.
%% The Lyapunov exponent test relies on the linearised map which
%% generally depends nonlinearly on the data. Moreover, the higher the
%% nonlinearity of the map, the higher the nonlinearity of the
%% Jacobian. In contrast, the $K$ test 
%% varies linearly with the data $\phi(n)$ {\em regardless}
%% of the form of the underlying dynamics. 
%% 
Moreover, in the case of high dimensional ODE's we found that our
method compares well also with the direct method proposed by Rosenstein
{\it et al.}~\cite{Rosenstein93} in the noise-free case, and shows much
better results in the case of underlying quasiperiodic dynamics in
the presence of measurement noise. 
Our test works well as a relative test and also as an absolute
test for moderately short time series contaminated by
measurement noise. 

Since our test applies directly to the time-series data, bypassing
the need for phase space reconstruction, we anticipate that the advantages
of our method will be magnified for data from partial differential equations
and from real experiments.   (However, the comparison is complicated by
the increased difficulty in generating the data and the fact that often
in these situations the ``correct'' answer is not known beforehand,
which is why a reliable test is required in the first place.)
This will be the subject of future work.

\paragraph{Acknowledgements}  
Both authors are grateful to Leonard Smith for helpful suggestions.
The research of GG was supported in part by the Australian 
Research Council.
The research of GG and IM was supported in part by 
EPSRC Grant GR/S22714/01.

GG acknowledges the hospitality of the Department of
Mathematics and Statistics of the University of Surrey where parts of
this work was done. 
IM is greatly indebted to the University of Houston for the use of e-mail,
given that pine is currently not supported on the University of Surrey network.

\end{document}